\documentclass[10pt,aps, prb , superscriptaddress, showpacs, floatfix, nofootinbib, svgnames, eqsecnum, eprint]{revtex4-2}

\usepackage{graphicx}
\usepackage {physics}
\usepackage{tikz}
\usetikzlibrary{shapes.geometric} 
\usepackage{footnote}
\usepackage{hyperref}
\usepackage[normalem]{ulem}
\usepackage{placeins}
\hypersetup{
    colorlinks = true,
    linkcolor = Red,     
    urlcolor = Green ,
    citecolor = Red ,
    }
    
\usepackage{bm}
\usepackage{xcolor}
\usepackage[caption = false]{subfig}
\usepackage{soul}
\begin{document}

\title{Intricately Entangled Spin and Charge Diffusion and the Coherence-Incoherence Crossover in the High-Dimensional Hubbard Model}
%\textcolor{green}{why not remove 'Spectral Analysis of'? } 
\author {Gopal Prakash}
\email {gopalp@imsc.res.in}
\affiliation {Institute of Mathematical Sciences, CIT Campus, Tharamani, Chennai 600113, India}
\affiliation {Homi Bhabha National Institute, Training School Complex, Anushakti Nagar, Mumbai
400085, India}

\author{S.R.Hassan}
\email {shassan@imsc.res.in}
\affiliation {Institute of Mathematical Sciences, CIT Campus, Tharamani, Chennai 600113, India}
\affiliation {Homi Bhabha National Institute, Training School Complex, Anushakti Nagar, Mumbai
400085, India}

\author{M.S. Laad}
\email {mslaad@imsc.res.in}
\affiliation {Institute of Mathematical Sciences, CIT Campus, Tharamani, Chennai 600113, India}
\affiliation {Homi Bhabha National Institute, Training School Complex, Anushakti Nagar, Mumbai
400085, India}

\author{N.S.Vidhyadhiraja}
\email {raja@jncasr.ac.in}
\affiliation{JNCASR, Jakkur P.O., Bangalore, India}

\author{T.V.Ramakrishnan}
\email {tvrama2002@yahoo.co.in}
\affiliation {Department of Physics, Indian Institute of Science, Bangalore, India}
\affiliation{JNCASR, Jakkur P.O., Bangalore, India}

\begin{abstract}
Correlation-driven metal-insulator transitions and temperature-driven quantum-coherent-to-incoherent crossovers in correlated electron systems underpin the doping, temperature and frequency-resolved evolution of physical responses. Motivated by recent experimental studies that investigate the evolution of dynamical spin and charge responses, we analyze the spin and charge diffusion spectra in both half-filled and doped one-band Hubbard model using Dynamical Mean Field Theory (DMFT) combined with the Numerical Renormalization Group (NRG). We compare the relative strengths and limitations of Density Matrix NRG (DMNRG) and Full Density Matrix NRG (FDM-NRG) in capturing low-frequency spectral features and their evolution with temperature, interaction strength and band-filling. Key measures, including characteristic frequency scales, Kullback-Leibler divergence, diffusion constants, and kurtosis provide complementary but internally consistent picture for the evolution of spin and charge excitations across bandwidth as well as the band-filling driven Mott transitions and the coherent-incoherent crossover. We find that spin and charge fluctuations cross over from quantum-coherent-to-quantum-incoherent at distinct temperatures, providing a microscopic insight into the complex, two-stage, Fermi-to-non-Fermi liquid-to-bad metal crossovers seen in transport data, in particular in the $dc$ resistivity. 
\end{abstract}

\maketitle

\section{Introduction}
The correlation-driven metal-insulator transition (MIT) \cite{NFMOTT,ImadaMott,PhysRevLett.121.067601,Charnukha,PhysRevLett.101.186403,Laad} remains of enduring interest in spite of extensive study over six decades. Though various aspects of bandwidth as well as filling-driven MITs \cite{Werner,Leshen,Kapcia} are well understood by now, the detailed evolution of the collective, dynamical charge and spin fluctuations remains a largely open issue of contemporary interest. Because magnetic fluctuations have been more extensively studied (inelastic neutron scattering and NMR) post the cuprate revolution in 1986, important progress in understanding of evolution of dynamical charge correlations across the MIT has only been achieved recently from careful resonant inelastic X-ray scattering (RIXS) \cite{PhysRevB.96.115148,suzuki2018probing,PhysRevB.98.161114,hepting2018three,PhysRevX.9.031042,doi:10.7566/JPSJ.88.075001,10.21468/SciPostPhys.3.4.026} and momentum-resolved electron energy loss spectroscopy (m-EELS)  \cite{10.21468/SciPostPhys.3.4.026,PhysRevX.9.041062}.  While the former measures coupled charge and spin correlations, the latter is dominantly a reflection of the dynamic dielectric function (more precisely, of $1/\epsilon({\bf q},\omega)$). These studies offer a motivation for detailed study of two-fermion dynamics, and a window to test various theoretical schemes. 

Another pertinent issue in this context relates to the evolution of dynamical spin and charge fluctuations across the quantum ``coherence-to-incoherence crossover'' that is often seen in a wide variety of correlated materials at low $T=T_{coh}$ near the MIT, (most notably in Sr$_{2}$RuO$_{4}$), where $T_{coh}\sim 20$~K and in DMFT and ECFL studies \cite{PhysRevLett.110.086401,Shastry1} of one- and multiband Hubbard models. Given that DMFT achieves remarkable (even quantitative) accord with a range of physical observables across \(T_{coh}\), one hopes that the same limit (where the irreducible vertices that enter the Bethe-Salpeter equations for the susceptibilities also become spatially local) will also offer a comparative accord with two-particle responses.  However, semi-analytic approaches are difficult to control while evaluating irreducible vertices, and hence, these issues have been sought to be tackled by numerical methods. Both continuous time quantum Monte Carlo (CTQMC) \cite{RevModPhys.83.349} and numerical renormalization group (NRG)  \cite{RevModPhys.47.773,RevModPhys.80.395} solvers have been employed, with their respective strengths and limitations (CTQMC is plagued by being too costly at very low \(T\), and by the mathematically ill-posed numerical analytic continuation, while NRG is hitherto mostly limited to the one-band Hubbard model). To our knowledge, NRG-based solvers have been scarcely used to study collective charge/spin dynamics.

In this work, we attempt to fill this gap.  We employ two variants of the NRG to investigate the evolution of charge and spin fluctuations across the MIT: \((i)\) we study their evolution across the MIT at very low \(T\), and \((ii)\) we study their evolution across the coherence-incoherence crossover mentioned above.  Armed with these, we discuss the implications of our findings for observations in extant data, as well as for the ubiquitous issue of the processes that lead to destabilization of the Landau quasiparticles across \(T_{coh}\). We conclude with broader implications of our results for systems where one-band modeling is insufficient.

In Section \ref{model_n_method}, we offer a concise overview of DMFT combined with NRG, outlining different schemes and defining the diffusion spectra. Section \ref{half_fill_hubb} features a comparative analysis of methods like FDM-NRG and DMNRG, emphasizing the challenges in accurately capturing low-frequency behavior. In Section \ref{doped_hubb}, we delve into the additional detailed information gleaned from various measures and their significance in understanding the system dynamics. The final section \ref{conclusion} presents our conclusions.

\section{Model and Method}
\label{model_n_method}
The one-band Hubbard Hamiltonian is

\begin{equation}
    \mathcal{H} = -t \sum_{\langle ij \rangle, \sigma} (c_{i\sigma}^{\dagger} c_{j\sigma} + \text{h.c.}) + U \sum_{i} n_{i\uparrow} n_{i\downarrow} - \mu \sum_{i\sigma} n_{i\sigma},
\end{equation}

where \( c_{i\sigma}^{\dagger} \) and \( c_{i\sigma} \) are the creation and annihilation operators for an electron with spin \( \sigma \) at site \( i \). The hopping amplitude between nearest-neighbor sites is represented by \( t \), and \( U \) indicates the on-site Coulomb repulsion. The electron number operator at site \( i \) with spin \( \sigma \) is \( n_{i\sigma} = c_{i\sigma}^{\dagger} c_{i\sigma} \), and \( \mu \) is the chemical potential. This Hamiltonian effectively captures the competition between electron delocalization, driven by kinetic energy, and localization, driven by on-site repulsion. The balance between these elements leads to various electronic phases such as metallic, Mott insulating, and correlated magnetic states, dependent on the ratio \( U/t \) and electron filling \( n \).

Dynamical Mean Field Theory (DMFT) \cite{georges1996} is employed to reduce the complexity of the lattice problem by mapping it onto a single-site impurity model embedded in a self-consistently determined electronic bath. The local Green's function \( G(\omega) \) is given by:
\begin{equation}
    G(\omega) = \int d\epsilon \, \rho(\epsilon) \left[ \omega + \mu - \epsilon - \Sigma(\omega) \right]^{-1},
\end{equation}
where \( \rho(\epsilon) \) is the non-interacting density of states, and \( \Sigma(\omega) \) represents the local self-energy derived from the impurity solution. The self-consistency condition in DMFT is achieved using the Weiss field (non-interacting impurity Green's function):
\begin{equation}
    \mathcal{G}_0^{-1}(\omega) = \omega + \mu - \Delta(\omega),
\end{equation}
with \( \Delta(\omega) \) denoting the hybridization function that characterizes the impurity-bath interaction. The iterative process refines \( \Sigma(\omega) \) and \( G(\omega) \) until convergence obtains, yielding an accurate portrayal of electronic correlation effects on spectral responses.  We have performed the calculations on a Bethe lattice where $\rho(\epsilon) = 2\sqrt{\epsilon^2 - D^{2}}/\pi D^2 $ and $\Delta(\omega) = t^2 G(\omega)$. Where $D = 2t$ is the half bandwidth.

To address the impurity model within DMFT, the Numerical Renormalization Group (NRG) is used, which excels at resolving systems with strong correlations. NRG discretizes the conduction band using logarithmic intervals and transforms the impurity model into a semi-infinite chain, with each subsequent site representing a progressively lower energy scale. Iterative diagonalization of the chain from high to low energy states enables precise determination of spectral properties near the Fermi level.

The NRG process involves band discretization by introducing a parameter \( \Lambda > 1 \), constructing a Wilson chain where energy scales at each site are proportional to \( \sim \Lambda^{-l/2} \) for $l$th site in the chain. By iteratively diagonalizing and retaining only low-energy states at each step, the method accurately calculates the Green's function and self-energy.

NRG facilitates the computation of local spectral functions, essential for understanding system behavior, such as spin and charge diffusion spectra. Analyzing dynamical correlations through the Green's function \( G_{AB}(t) \) for operators \( A \) and \( B \), which describes their time-dependent correlation, provides deeper insight:
\begin{align}
    G_{AB}(t) = -i \theta(t) \langle [A(t), B]_{\pm} \rangle,
\end{align}
where \( \pm \) indicates the anticommutator for fermionic operators and the commutator for bosonic ones. The thermal average \( \langle \cdots \rangle \) is taken in the grand canonical ensemble, ensuring accurate quantum statistics. The frequency representation of \( G_{AB}(t) \) is:
\begin{align}
    G_{AB}(\omega) = \int_{-\infty}^{\infty} \frac{\rho_{AB}(\omega_1)}{\omega - \omega_1} \, d\omega_1,
\end{align}
where the spectral density \( \rho_{AB}(\omega) \) is defined by:
\begin{align}
    \rho_{AB}(\omega) = -\frac{1}{\pi} \, \text{Im} \, G_{AB}(\omega).
\end{align}

The local Green's function for the impurity, with \( A \) and \( B \) as \( c \) and \( c^\dagger \), yields the density of states \( A(\omega) \):
\begin{align}
    A(\omega) = -\frac{1}{\pi} \mathrm{Im} G(\omega).
\end{align}

Dynamic susceptibilities for charge and spin, \( \chi_{n/s}(\omega) \), utilize the number operator \( n \) and spin operator \( s \):
\begin{align}
    n &= \sum_{\sigma = \uparrow,\downarrow} c^\dagger_{\sigma} c_{\sigma}, \\
    s &= c^{\dagger}_{\uparrow} c_{\uparrow} - c^{\dagger}_{\downarrow} c_{\downarrow},
\end{align}
and are expressed as:
\begin{align}
    \chi_{n}(\omega) &= \int_{-\infty}^{\infty} dt \, e^{i \omega t} \left( -i \theta(t) \langle [n(t), n]_{-} \rangle \right), \\
    \chi_{s}(\omega) &= \int_{-\infty}^{\infty} dt \, e^{i \omega t} \left( -i \theta(t) \langle [s(t), s]_{-} \rangle \right).
\end{align}

The spectral density \( \rho_{AB}(\omega) \) can be represented using the Lehmann representation:
\begin{align}
    \rho_{AB}(\omega) = \sum_{mn} e^{-\beta E_n} \langle n | A | m \rangle \langle m | B | n \rangle \, \delta(\omega + E_n - E_m) \, (1 \pm e^{-\beta \omega}),
\end{align}
with the correct statistical factors for fermionic (+1) and bosonic (-1) operators. Spectral function calculations employ Density Matrix NRG (DMNRG) \cite{hofstetter}  and Full Density Matrix NRG (FDM-NRG) \cite{FDMNRG}. The DMNRG method performs iterations until the desired temperature \( T_{N} \) is reached. The density matrix is defined by:
\begin{align}
\hat{\rho} = \sum_{m} e^{-\beta_{N}E^{N}_{m}} \ket{m}_{N} \bra{m}.
\end{align}

\begin{figure}[h]
    \centering
    \includegraphics[width=0.7\linewidth]{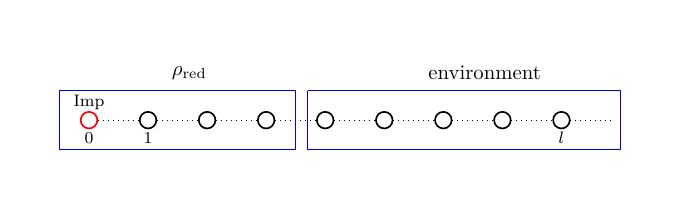}
    \caption{Reduced density matrix obtained by tracing out “environment” degrees of freedom of the Wilson chain.}
    \label{rdm}
\end{figure}

Each NRG step \( l \) provides the spectral density at \( \omega \sim T_{l} \). The chain, divided into a cluster of length \( l \) and an environment, as illustrated in Fig. \ref{rdm}, yields:
\begin{align}
\hat{\rho} = \sum_{m_{1} m_{2} n_{1} n_{2}} \rho_{m_{1} n_{1}, m_{2} n_{2}} \ket{m_{1}}_{env} \ket{n_{1}}_{sys} \bra{n_{2}} \bra{m_{2}}.
\end{align}

The reduced density matrix, obtained by tracing out the environment, is:
\begin{align}
\hat{\rho}^{red} = \sum_{n_{1} n_{2}} \rho^{red}_{n_{1} n_{2}} \ket{n_{1}}_{sys} \bra{n_{2}},
\end{align}
with:
\begin{align}
\rho^{red}_{n_{1} n_{2}} = \sum_{m} \rho_{m n_{1}, m n_{2}}.
\end{align}

The FDM-NRG \cite{FDMNRG} method constructs a full density matrix using the complete Anders Schiller (AS) basis \cite{PhysRevLett.95.196801}, ensuring sum rules are preserved:
\begin{align}
\hat{\rho} = \frac{e^{-\beta \hat{H}}}{Z} = \sum_{l} w_l \hat{\rho}_{l}^{\mathrm{DD}},
\end{align}
where \( \hat{\rho}_{l}^{\mathrm{DD}} \) represents the partial density matrix for discarded states of shell \( l \), and \( w_l \) distributes contributions by thermal significance:
\begin{align}
A_{BC}(\omega) = \sum_{l} w_l A_{BC}^{l}(\omega),
\end{align}
with \( A_{BC}^{l}(\omega) \) being the shell \( l \) contribution.

To obtain the continuous spectra from the discrete NRG energy levels, We use spectral broadening. Logarithmic discretization can cause low-frequency oscillations. To smooth results, we apply a broadening kernel transitioning from log-Gaussian to Gaussian:
\begin{equation}
\begin{aligned}
    K(\omega, E) &= \frac{\theta(\omega E)}{\alpha |\omega| \sqrt{\pi}} \exp\left(-\left(\frac{\log(\omega/E)}{\alpha} - \gamma\right)^2\right) h(|\omega|) \\
    &+ \frac{1}{\omega_{0} \sqrt{\pi}} \exp\left(-\left(\frac{\omega - E}{\omega_{0}}\right)^2\right) \left(1 - h(|\omega|)\right), \\
    h(\omega) &= \exp\left(-\left(\frac{\log(\omega/\omega_{0})}{\alpha}\right)^2\right), \quad \gamma = \alpha/4,
\end{aligned}    
\end{equation}

with parameters \( \alpha = 0.5 \) and \( \omega_{0} = 10^{-99} \). The kernel provides smooth spectral functions while retaining key details. Calculations used NRG Ljubljana \cite{NRGLJ} with \( \Lambda = 2 \), \( N_z = 8 \), and \( N_{\text{keep}} = 10000 \). Enhanced convergence was ensured with a refined self-energy method \cite{new_self_energy_trick}.

Charge and spin spectral functions were obtained using DMNRG and FDM-NRG, with analytic continuation via Pade approximants \cite{Zitko_Pade} applied to FDM data. The diffusion spectrum \( P_{n/s}(\omega) \) \cite{PhysRevB.110.075106}  was analyzed to understand dissipation mechanisms:
\begin{align}
    P_{n/s}(\omega) = \frac{\Im \chi_{n/s}(\omega)}{\pi \omega \chi_{n/s}},
\end{align}
revealing energy dissipation characteristics, coherence-incoherence transitions, and their implications for transport properties in correlated systems. All the energy scales in the subsequent sections are presented in terms of half band width $D$. So that the $\omega$ denotes $\omega/D$.

\FloatBarrier

\section{Half Filled Hubbard Model}
\label{half_fill_hubb}
\subsection{Comparative Analysis of Diffusion Spectra Using DMNRG and FDM Methods}

In this subsection, we delve into a comparative analysis of the computed charge and spin diffusion spectra, \( P_{n/s}(\omega) \), derived from two computational techniques: the Density Matrix Numerical Renormalization Group (DMNRG) and the Full Density Matrix (FDM) methods. Each approach has its strengths and limitations. DMNRG is particularly effective for low-temperature analyses, while FDM is more suitable for intermediate to high-temperature ranges. For this study, we focus on half-filling calculations at a low temperature \( T/D = 0.001 \), comparing the effectiveness and accuracy of each method in capturing low-frequency behavior and overall spectral features.

\begin{figure}
    \centering
    \subfloat[]{\includegraphics[width=0.5\linewidth]{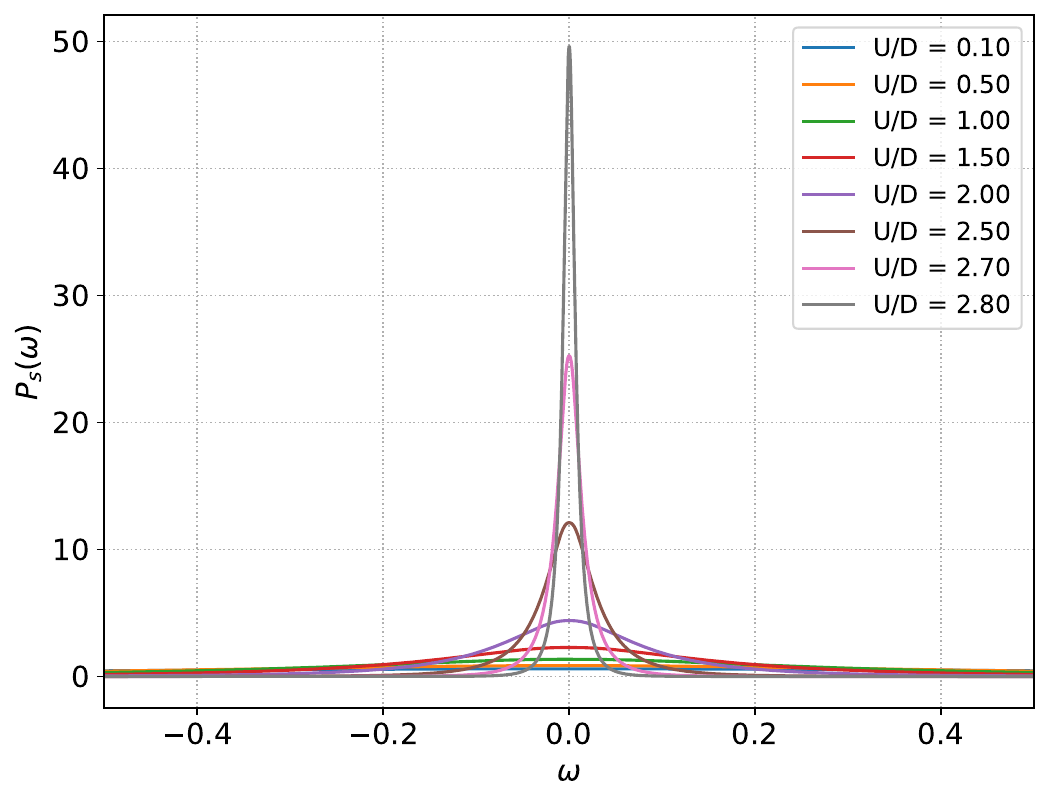}}
    \subfloat[]{\includegraphics[width=0.5\linewidth]{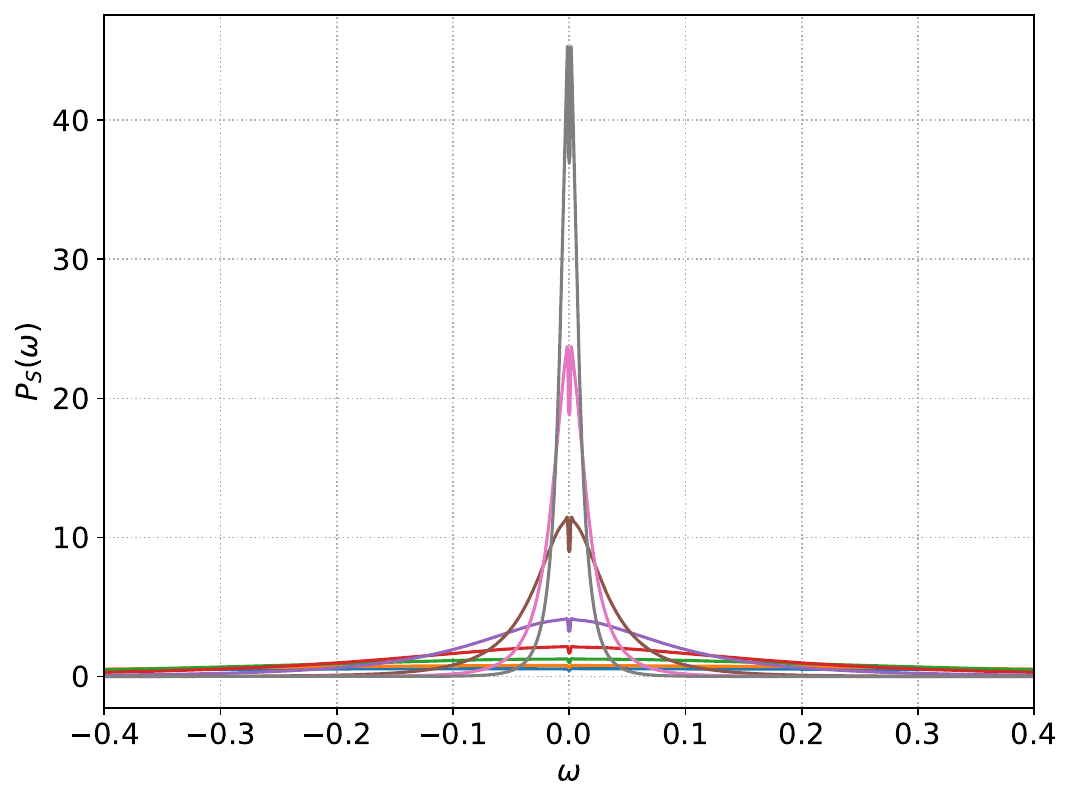}}
    \caption{(a) Spin diffusion spectrum for half-filling using DMNRG at $T/D = 0.001$. (b) Spin diffusion spectrum for half-filling using FDM at $T/D = 0.001$.}
    \label{spindiff_hf}
\end{figure}

In Fig. \ref{spindiff_hf}, we show the evolution of the spin diffusion spectrum (SDS) as a function of \( U/D \) for the half-filled Hubbard model at very low temperature, \( T/D = 0.001 \), for the metallic phase (\( U/D \leq 2.85 \)). It is well known that a weakly correlated itinerant Landau FL (LFL) metal smoothly evolves into a strongly correlated LFL metal as \( U/D \) increases. The latter is marked by a narrow ``Lattice Kondo resonance" and corresponds to coherent propagation of heavily renormalized Landau quasiparticles. This is associated with the self-consistent screening of local magnetic moments via a lattice Kondo effect. Above \( T_{coh} \), thermal unbinding of the local Kondo singlet reinstates strong local moment fluctuations, quenching the Kondo resonance.

Mirroring this evolution, the SDS function, \( P_s(\omega) \), smoothly evolves from a broad (at small \( U/D \)) to a progressively narrower lineshape centered symmetrically around \( \omega = 0 \) at large \( U/D \). We interpret this evolution as follows: at small \( U/D \), well-defined local moments cannot form because they are washed out by strong itinerance. Near the MIT, progressively stronger local moment formation and fluctuation tendencies obtain due to strongly reduced itinerance (it must be borne in mind that the reduction of itinerance and tendency to enhanced local moment fluctuations arise from one single aspect, i.e., dynamical competition between \( U \) and \( t \). The enhanced local moment fluctuations manifest as strongly enhanced low-energy spin fluctuations, and this is directly reflected in the enhanced \( P_{s}(\omega) \). Ultimately, in the Mott insulator (\( U/D > 3.0 \)), \( P_{s}(\omega) \) collapses into a delta function peak at \( \omega = 0 \). This is simply a reflection of the spatially uncorrelated local moment fluctuations of the paramagnetic Mott insulator that give an extensive ground state degeneracy (\( O(\ln2) \) per site), and is specific to DMFT. Beyond DMFT, such local moments would form spin singlets at low energy and quench these features.  Since charge fluctuations are energetically costly in the Mott insulator, this is a (trivial) manifestation of spin-charge separation.

We now compare how DMNRG fares relative to FDM. The DMNRG-derived spectrum, depicted in Fig. \ref{spindiff_hf}(a), reveals notable changes as the interaction strength \( U \) increases. For small \( U \), the spectrum near \(\omega = 0\) displays a broad structure, indicating weaker correlations and more delocalized spin excitations. However, as \( U \) approaches the Mott transition, the spectrum sharpens significantly, forming a pronounced peak at \(\omega = 0\).  The attractiveness of the DMNRG is it's ability to yield a continuous lineshape: this desirable feature is not shared by the FDM, as we discuss below.

The FDM results, shown in Fig. \ref{spindiff_hf}(b), present a similar overall trend but with distinct deviations.  An unexpected dip appears at \(\omega = 0\), deviating from the expected physical behavior. This anomalous feature is likely a numerical artifact inherent to the FDM approach. It indicates that FDM cannot accurately represent the behavior of the spin susceptibility at low frequencies. This slope error, when divided by \(\omega\) to calculate the diffusion spectrum, manifests as a pronounced spurious dip at low frequencies.

The appearance of this artifact underscores the limitations of the FDM method in resolving low-frequency behaviors compared to the DMNRG method, which maintains a more consistent and physically accurate profile. Understanding these discrepancies is crucial for assessing the reliability and applicability of each method in studying spin dynamics, particularly in systems where low-frequency precision is essential.

\FloatBarrier

\subsection{Analysis of the Charge Diffusion Spectrum}

\begin{figure}[h]
    \centering
    \subfloat[]{\includegraphics[width=0.5\linewidth]{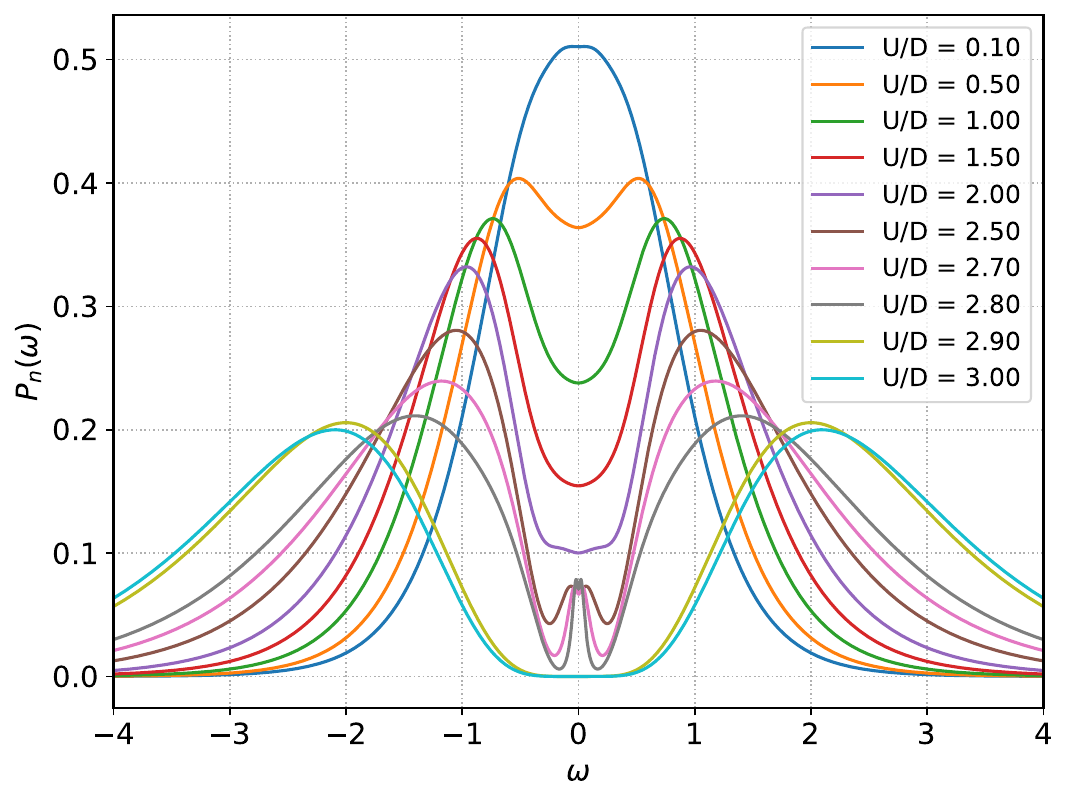}}
    \subfloat[]{\includegraphics[width=0.5\linewidth]{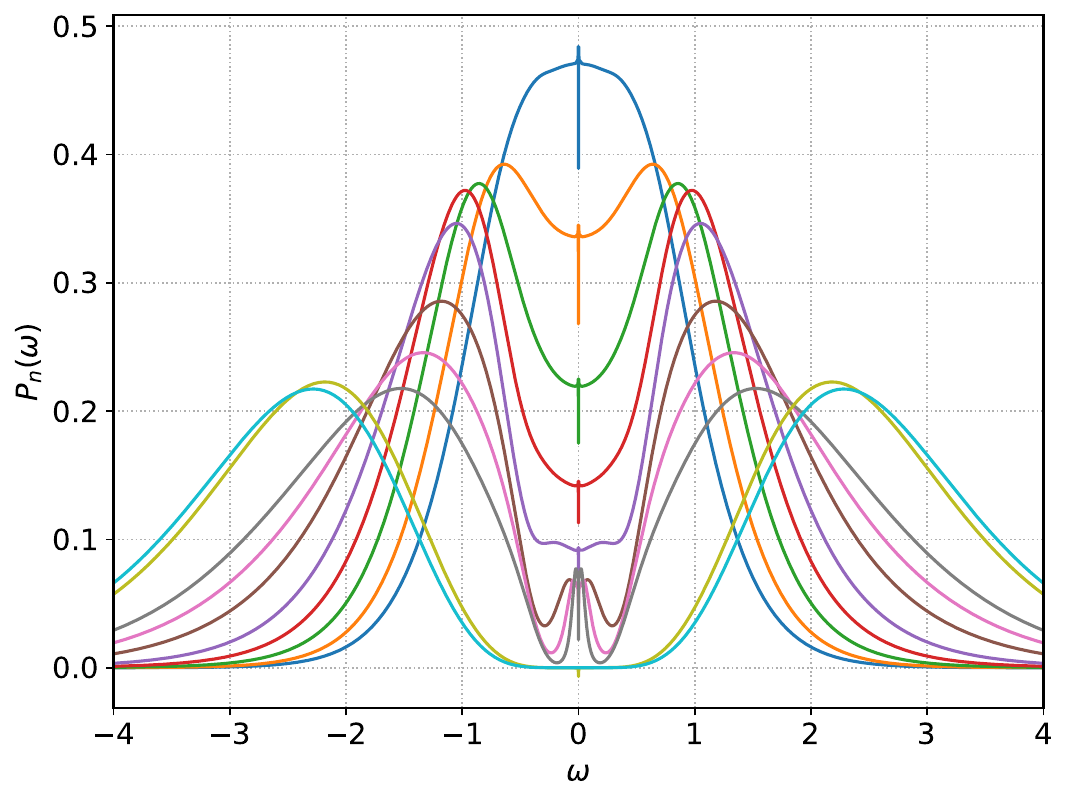}}
    \caption{(a) Charge diffusion spectrum for half-filling using DMNRG at $T/D = 0.001$. (b) Charge diffusion spectrum for half-filling using FDM at $T/D = 0.001$.}
    \label{chargediff_hf}
\end{figure}

The charge diffusion spectrum, shown in Fig. \ref{chargediff_hf}, exhibits very different features. For small-to-moderate \( U/D \), the low \( T/D = 0.001 \) CDS for half-filling features a two sub-band structure that ``spreads out" with a continuous low-energy reduction with increasing \( U/D \). In the strongly correlated FL regime, however, a new feature  obtains: \( P_{n}(\omega) \) now develops a drastically narrowed low-energy quasicoherent split-peak feature, coexisting with high-energy incoherent bands.  Interestingly, the height of this peak remains almost fixed, and it's width small, throughout the regime where the one-fermion spectrum exhibits a sharp lattice Kondo resonance (slightly broadened above $T_{coh}$ and very narrow below it, in the true Landau FL metal) coexisting with incoherent broad bands.

We interpret this evolution of $P_{n}(\omega)$ as follows: it is known, though not sufficiently appreciated, that the true symmetry group of the half-filled Hubbard model is \( O(4)/Z_{2}=SU(2)\times SU(2) \), associated with separately conserved spin and charge currents \cite{anderson2001symmetries}. While this is well known since 1990~\cite{Yang-IJMPB}, its manifestations in the evolution of charge and spin fluctuations in the Hubbard model are not documented clearly, to our best knowledge. 

Given charge pseudospins, $\tau^{+} = c_{\uparrow}^{\dagger}c_{\downarrow}^{\dagger}, \tau^{-} = c_{\downarrow}c_{\uparrow}, \tau^{z} = (1-n_{\uparrow}-n_{\downarrow})/2$ satisfying the $SU(2)$ symmetry, and related to the spin-$1/2$ operators by a particle-hole transformation in one, here $\downarrow$ spin species, one would naturally expect one-fermion hopping to cause charge-pseudospin fluctuations as well. At small $U/D$, charge-pseudospin formation is prevented by large $t$, and thus, these have no time to form and coherently propagate. This is mirrored in the broad two-band structure seen in $P_{n}(\omega)$. In the strongly correlated metal, however, larger $U/t$ leads to heavy, hence slowly propagating quasiparticles in the infrared. This allows charge pseudospins to ``preform" and, depending upon $t$, to undergo a separate ``Kondo-like" quenching. Clearly, this latter screening can only occur when carriers with drastically reduced kinetic energy have first formed: this does not require the usual Kondo screening to occur, but merely that the carriers have become ``heavy'' enough to permit ``preformed'' charge pseudospins.  In other words, this requires that the charge pseudospin fluctuation time scale, $\tau_{ch}\simeq \hbar/k_{B}T_{coh}^{ch}$ become comparable to that of the single fermion hopping, $\tau\simeq \hbar/D$. 

\begin{figure}[h]
    \centering
    \includegraphics[width=0.6\linewidth]{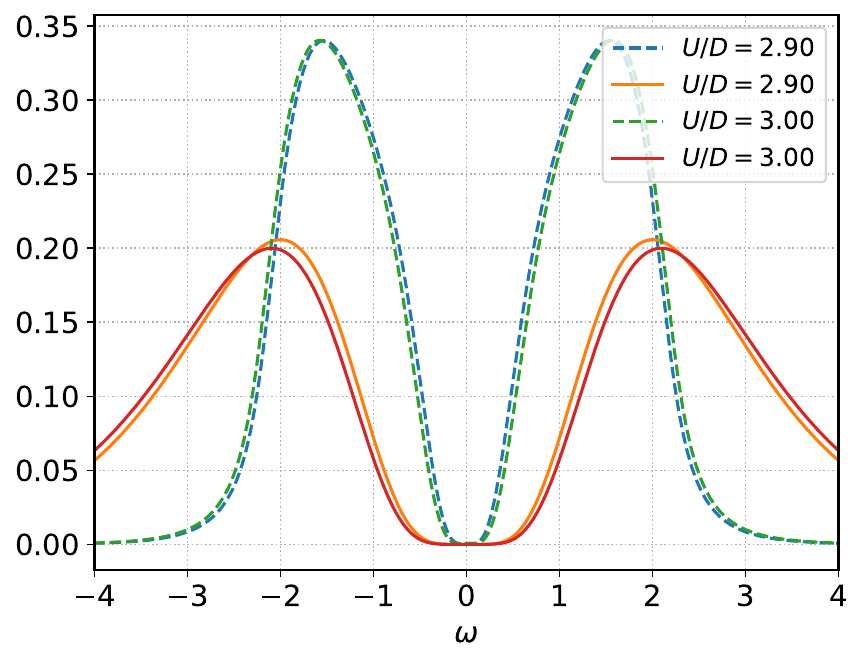}
    \caption{Energy gap comparison for single-particle DOS and charge diffusion spectra. Dotted lines represent the single-particle density of states, and solid lines represent the charge diffusion.}
    \label{energy_gap}
\end{figure}

Eventual emergence of the true Landau FL metal necessarily requires complete quenching of both, spin and charge pseudospin fluctuations. This implies two separate scales at which the spin (\( T_{coh}^{s} \)) and charge (\( T_{coh}^{ch} \)) fluctuations become quantum coherent, with the eventual FL scale set by \( T_{FL} = T_{coh}^{s} < T_{coh}^{ch} \). This two-stage quenching of incoherence must show up in physical observables. It is well known, for example, that the \( T \)-dependent DC resistivity exhibits a ``high-\( T \)" bad metal regime, where \( \rho_{dc}(T) \simeq \rho_{0} + A_{1}T \) exceeding the Mott-Ioffe-Regel (MIR) limit, a quantum-incoherent regime where \( \rho_{dc}(T) \simeq A_{2}T \) is less than the MIR limit, and an eventual low-\( T \) regime below \( T_{FL} \simeq 0.002-0.003 \) where \( \rho_{dc}(T) \simeq BT^{2} \).  It is now tempting to associate these three regimes with our findings: the bad metal is one where neither spin nor charge fluctuations are quenched, the quantum-incoherent metal with a regime where charge fluctuations are quenched but spin fluctuations are not, and the true FL metal with a regime where both are quenched. It should be interesting to seek to quantitatively match these regimes to the observed crossovers in the resistivity. In the insulating phase, at \( U/D = 0.29 \) and \( 0.30 \), the charge diffusion spectrum near \(\omega = 0\)  a clear charge gap characteristic of the Mott insulator.  Low-frequency charge excitations are energetically unfavorable. The observed charge gap is approximately twice the single-particle density of states gap (Fig. \ref{energy_gap}), signifying the combined energy required for particle-hole excitations. 

Next, we compare the results from DMNRG and FDM as shown in Fig. \ref{chargediff_hf}(a) and \ref{chargediff_hf}(b). Again, DMNRG produces a reliable profile with minimal spurious features, particularly at low \(\omega\) values, making it a preferred approach for capturing accurate low-frequency behavior in the charge diffusion spectrum. In the Kondo regime, the DMNRG results show a subtle peak near \(\omega = 0\), reflecting strong correlations where charge fluctuations are moderately suppressed due to the formation of a Kondo resonance. Beyond this regime, any artifacts present in DMNRG results are minimal and do not significantly affect the interpretation of charge diffusion data.

As expected, FDM shows limitations in capturing low-frequency features. As shown in Fig. \ref{chargediff_hf}(b), a persistent spurious dip appears at \(\omega = 0\) across all tested values of \( U \). Thus, FDM presents problems in accurately studying the charge susceptibility at low frequency. The slope error becomes amplified when normalized by \(\omega\), resulting in the observed dip. In Appendix \ref{app_A} we have presented a detailed analysis on the low frequency slope of charge and spin spectral function and in Appendix \ref{app_B} we explore strategies for minimizing these artifacts within the FDM approach, aiming for improved low-frequency precision and reliability.

The underlying cause of such artifacts in the FDM approach may stem from its discretization scheme or challenges associated with numerical continuation techniques used to extrapolate data from imaginary to real frequencies. While FDM is valuable for broader spectrum analyses, these findings emphasize its limitations in providing the same level of detail as DMNRG, particularly for low-frequency physics where precision is critical. This comparison, as illustrated in Fig. \ref{chargediff_hf}, shows that DMNRG yields a smoother, more consistent spectrum with minimal noise, whereas FDM, despite capturing the overall trend, requires cautious interpretation near \(\omega = 0\).

\FloatBarrier

\section{Doped Hubbard Model}
\label{doped_hubb}
 
\begin{figure}
    \centering
    \subfloat[]{\includegraphics[width=0.45\linewidth]{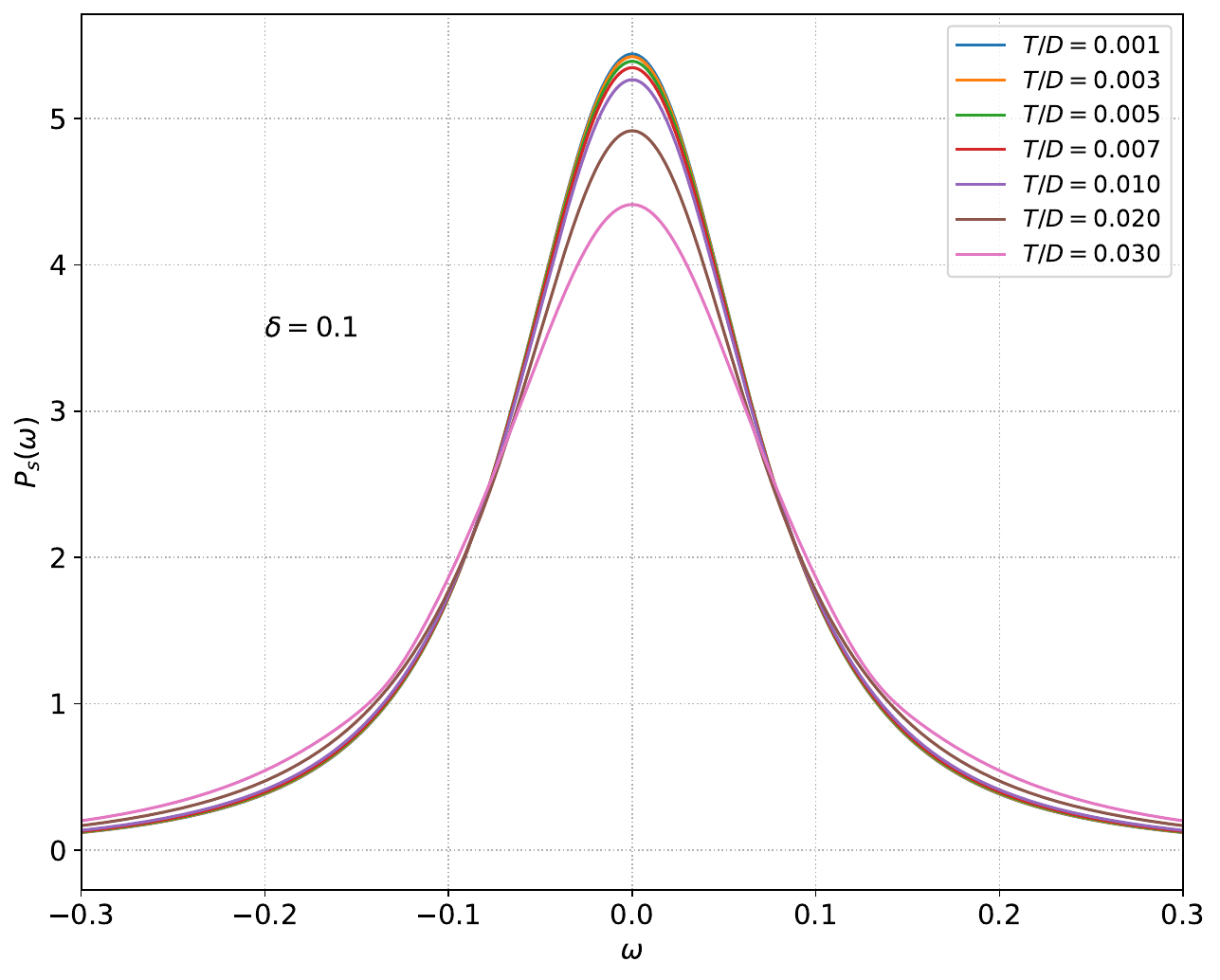}}
    \subfloat[]{\includegraphics[width=0.45\linewidth]{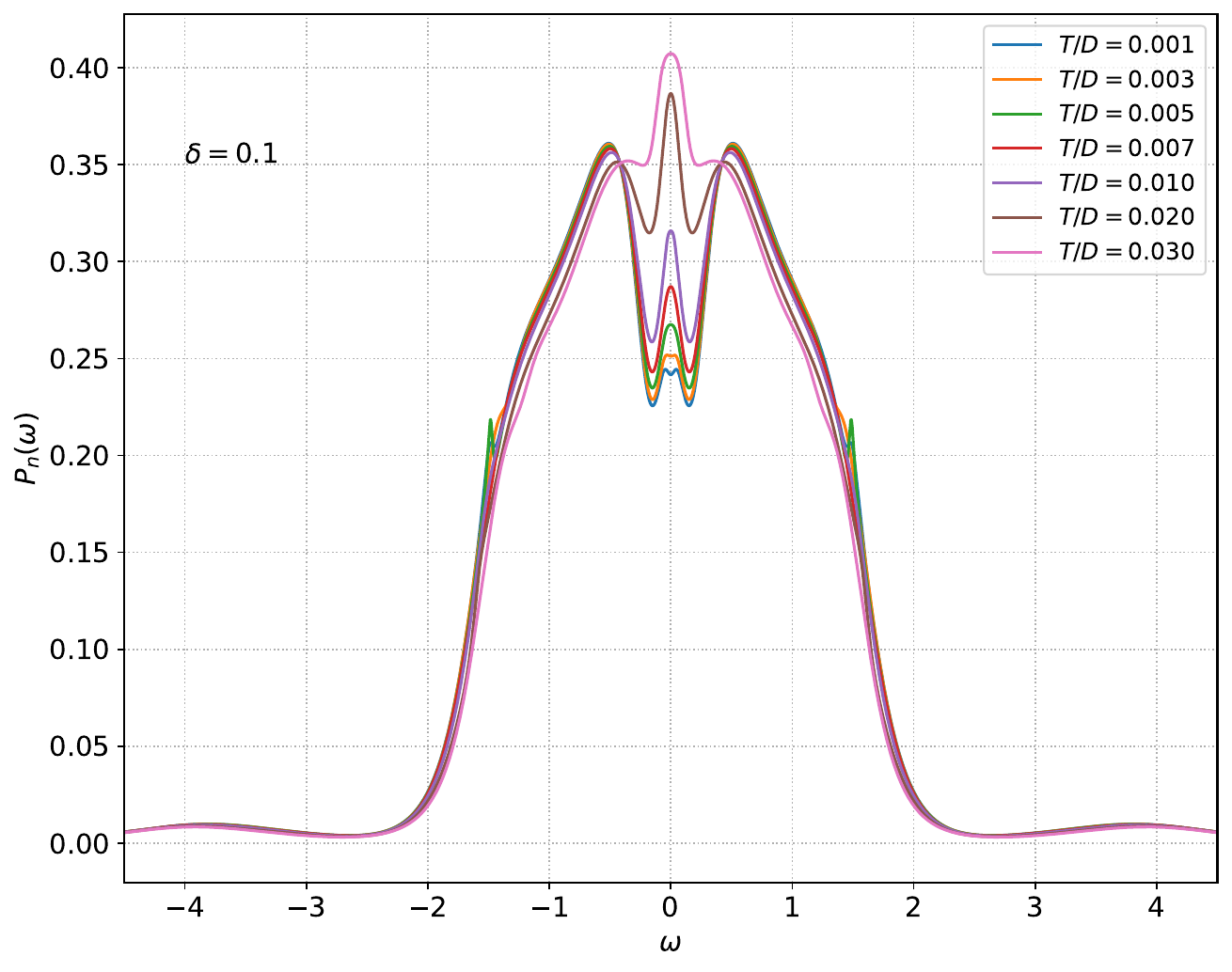}}
    \caption{(a) Spin diffusion spectrum and (b) charge diffusion spectrum at various temperatures for \(\delta = 0.1\) obtained through Pade analytic continuation (\(\eta/D = 0.05\)).}
    \label{dop_diff_spec_spin_charge}
\end{figure}

We now explore the spin and charge diffusion spectra for a doped Mott insulator with \( U = 4.0 \), examining doping levels of \( \delta = 0.1, 0.15, 0.2, \) and \( 0.25 \). Figure \ref{dop_diff_spec_spin_charge} illustrates the computed spectra at \( \delta = 0.1 \). These results were derived through Pade analytic continuation applied to Full Density Matrix NRG (FDM-NRG) data, using an optimized parameter value of \(\eta = 0.05\) to ensure reliability.

\begin{figure}
    \centering
    \includegraphics[width=0.65\linewidth]{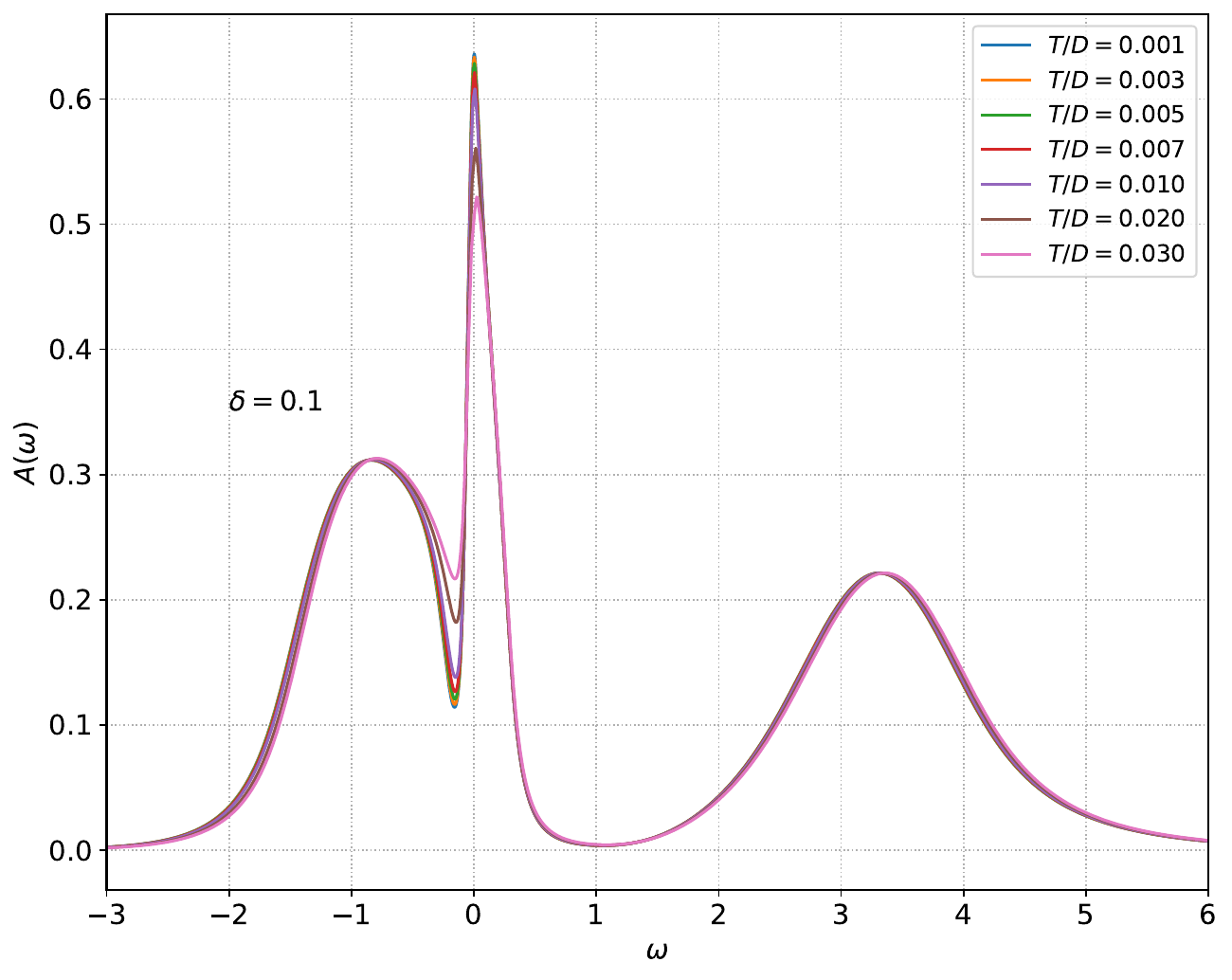}
    \caption{Density of states at various temperatures for \(\delta = 0.10\).}
    \label{DOS}
\end{figure}

The spin diffusion spectrum is characterized by a prominent peak at \(\omega = 0\), stemming from low-energy spin excitations. In contrast, the charge diffusion spectrum displays a more complex structure, with distinct peaks at both high and low energies. For comparison, Fig.\ref{DOS} illustrates the single-particle electron spectral function, which reveals a clear FL quasiparticle peak at comparable doping levels.  A notable feature in both spin and charge spectra is the clear appearance of isosbestic points, {\it i.e}, energy scales at which the responses are independent of $T$.  For the SDS, this occurs at $\omega_{s}\simeq 0.07$ while for the CDS, the isosbestic point is at $\omega_{c}\simeq 0.5$.  A direct comparison with the spectra at half-filling shows that the SDS retains the features of the undoped system apart from increased spin fluctuation damping due to enhanced carrier mobility.  On the other hand, the CDS features clear, new low-energy structures relative to those for the undoped case.

Away from half-filling, the chemical potential term, $\mu\sum_{i}n_{i}$ with $n_{i}=\sum_{\sigma}c_{i\sigma}^{\dagger}c_{i\sigma}$ directly couples to charge but not spin.  It acts like an ``external field'' coupled to charge fluctuations, leading to finite charge compressibility but leaving the spin fluctuations unaffected, except to the extent that the latter are selfconsistently modified by the charge fluctuation-modified fermion dynamics.  This breaks the SU$(2)$$\times$SU$(2)$ symmetry of the half-filled Hubbard model down to U$(1)$$\times$SU$(2)$.  Hence, with a non-zero $\mu$, violent phase fluctuations in the Mott insulator must be suppressed, translating into more quasi-coherent charge dynamics at low energy.  This translates into a low-energy feature (in the Kondo regime at large $U/D$) in the CDS that is enhanced relative to it's undoped value, but it's form at low $T$ reains similar.  However, as $T$ increases, spectral weight transfer from high- to low energy yields an overdamped peak centered around $\omega=0$, and reflects the thermal destruction of the charge pseudospin quenching.
  
The reduction of the weight of the low-energy feature relative to the high energy incoherent structures is again, as in the undoped case, a reflection of the fact that one-fermion excitations must first be ``well-formed'' to give rise to coherent low-energy charge fluctuations.  As $T$ increases, dramatic transfer of dynamical spectral weight from high- to low energy is clearly manifest, as is the observation that this is completely incoherent.  In the SDS, increasing $T$ damps out the central peak, transferring coherent weight across the isosbestic point, to high-energy incoherent regions.  The origin of these isosbestic points can, we believe, be traced back to sum rules involving charge and spin susceptibilities, but we have not done this here.  It is known that isosbestic points in the one-fermion spectral function of the Hubbard model within DMFT are a consequence of Luttinger's sum rule and dynamical spectral weight transfer \cite{eckstein2007isosbestic}. It would be interesting to be able to link the isosbestic points we find to the compressibility sum rule (integral of the charge susceptibility) and the local moment sum rule (integral of the local dynamical spin susceptibility) along with dynamical spectral weight transfer in the particle-hole sector.  We defer this issue for the future.

 We build upon and extend these findings by utilizing measures like characteristic frequency scales, Kullback-Leibler deviation (KLD), kurtosis, and the diffusion constant. Each metric offers unique insights into the spin and charge excitation characteristics, enabling a detailed understanding of their spectral dynamics. Characteristic frequency scales denote the average energy scale of excitations, whereas KLD measures deviations from reference distributions, thereby identifying subtle spectral variations. Kurtosis examines the spectral shape and tail, revealing sharp or diffuse features. The diffusion constant relates to transport properties, showing how well spin and charge excitations propagate within the system. Collectively, these tools provide a comprehensive exploration of transitions from coherent to incoherent regimes and the intricate relationship between spin and charge dynamics in strongly correlated systems.

\FloatBarrier

\subsection{Characteristic Frequency Scale and Quantum-Classical Transitions}
The diffusion spectrum \( P_{n/s}(\omega) \) is defined for both positive and negative frequencies, capturing all energy dissipation processes, including absorption and emission of local, dynamical charge and spin ``bosons'' by a propagating fermion in the medium. However, the characteristic frequency scale \( \Omega_{n/s} \) is defined only over positive frequencies:
\[
\Omega_{n/s} = \int_{0}^{\infty} d\omega \, \omega \, P_{n/s}(\omega).
\]

and typifies the energy scale of excitations and their dissipation within the system.  It thus arises as the mean energy scale necessary for distinct regimes of fluctuation-dominated transport. Integrating only positive frequencies highlights energy absorption and excitation processes, offering a clearer understanding of how energy is distributed and how the system transitions from quantum to classical behavior. By comparing \( \Omega_{n/s} \) to the thermal energy \( k_B T \), we can assess whether the behavior is quantum or classical.

\begin{figure}[h]
    \centering
    \includegraphics[width = 0.8 \linewidth]{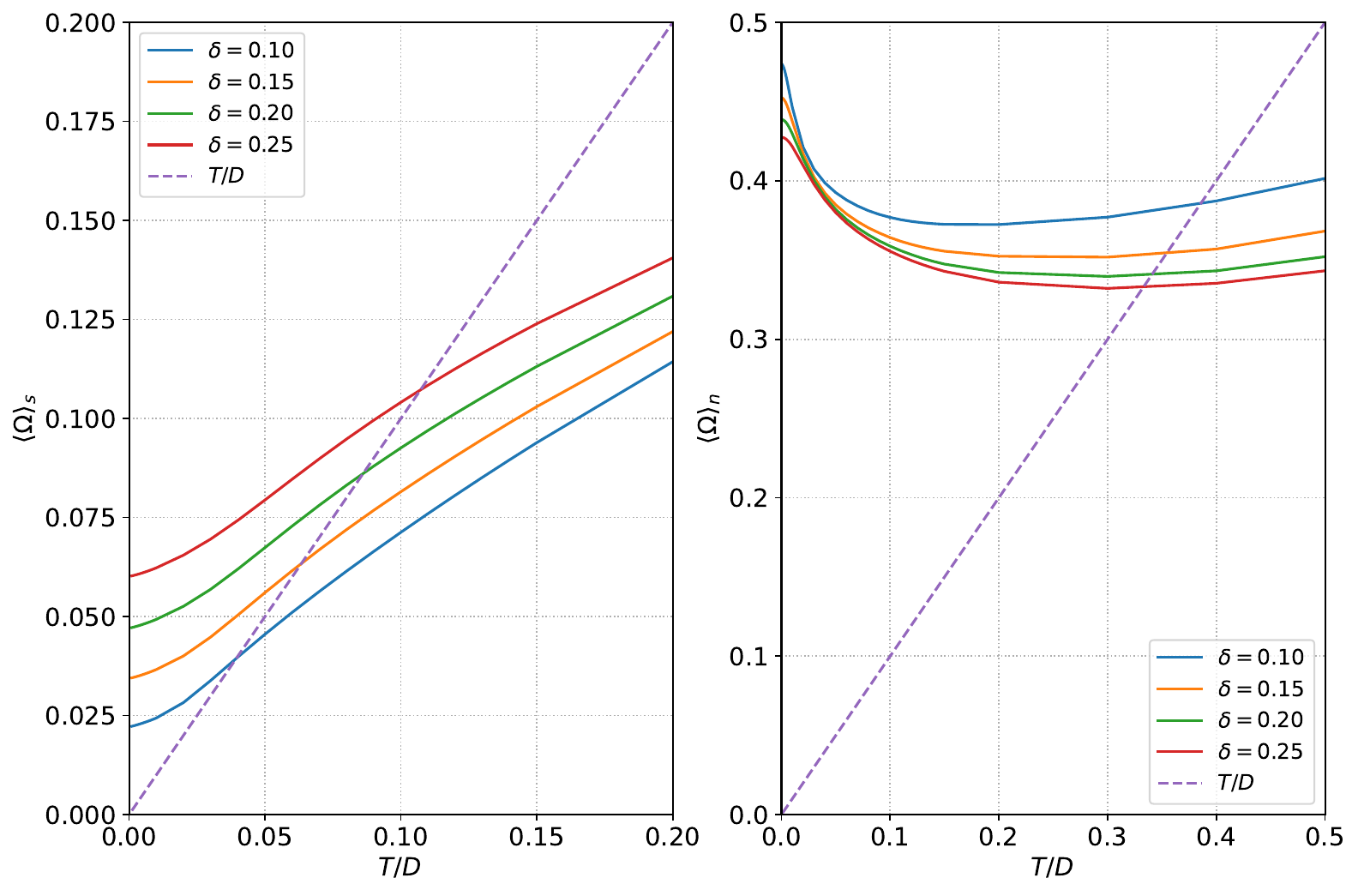}
    \caption{spin and charge average frequency of fluctuation at $U=4$ for different dopings $\delta = 0.1,0.15,0.2,0.25$.}
    \label{characteristic_freq_scale}
\end{figure}

\begin{figure}[h]
    \centering
    \subfloat[]{\includegraphics[width=0.45\linewidth]{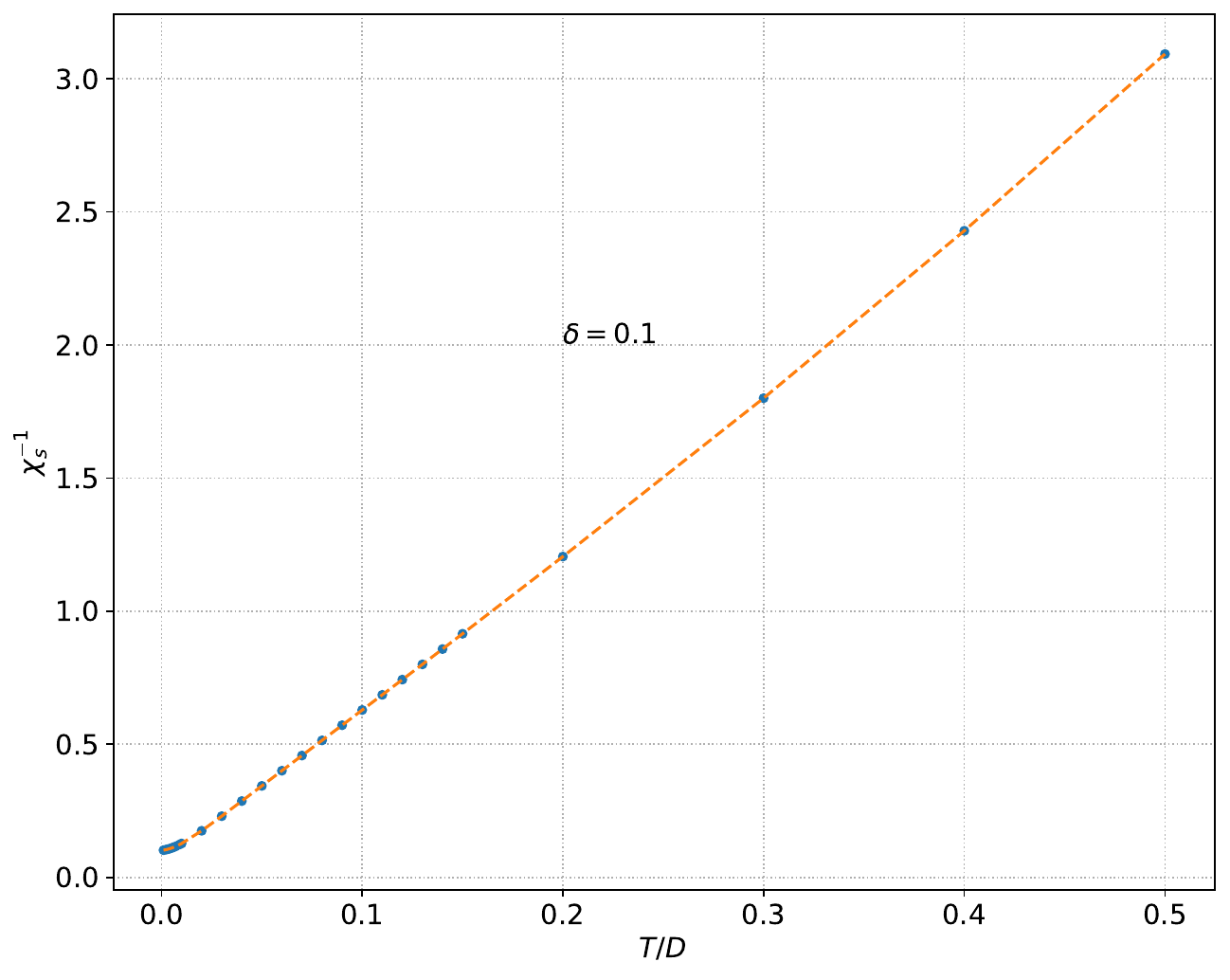}}
    \subfloat[]{\includegraphics[width=0.45\linewidth]{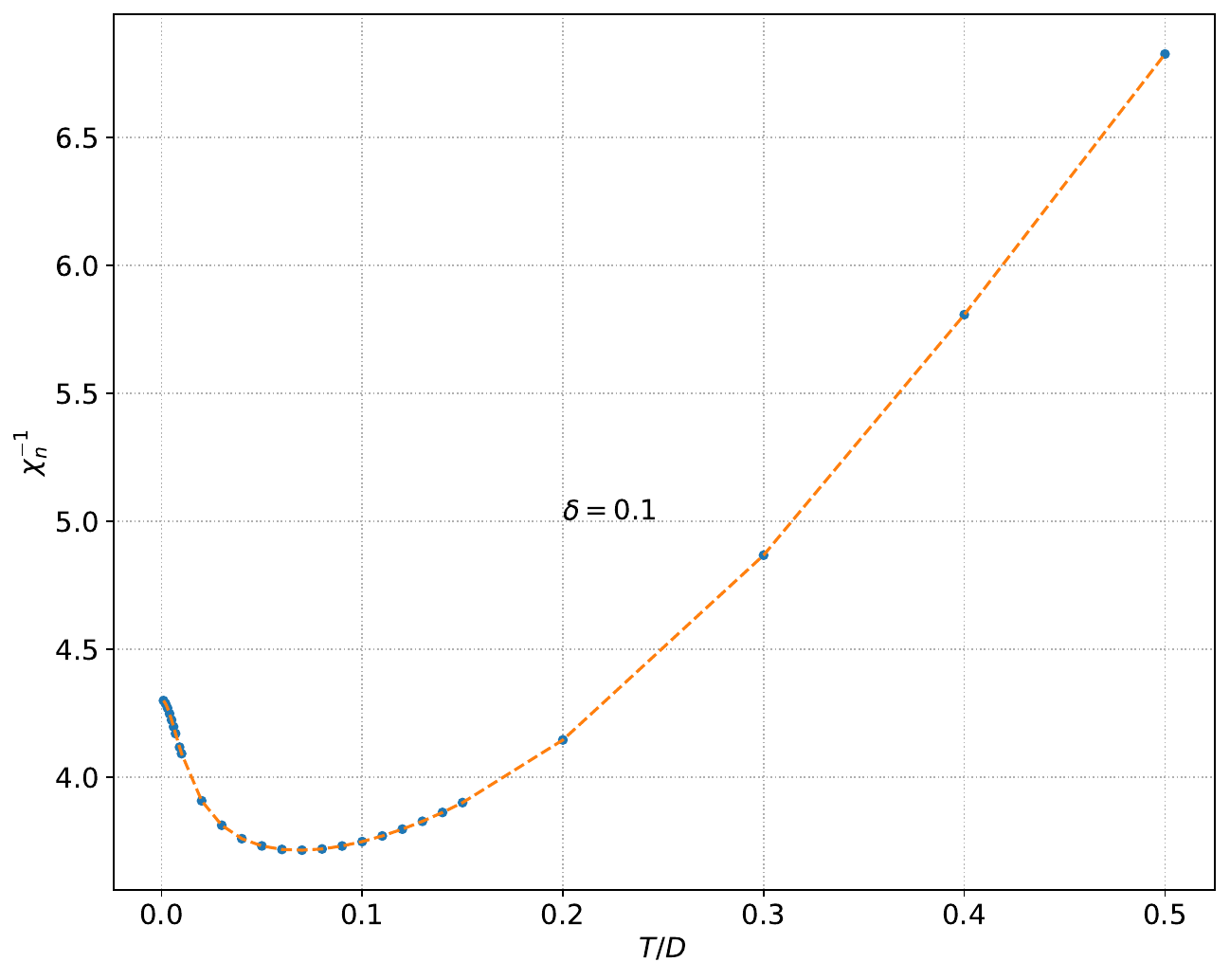}}
    \caption{(a) inverse spin susceptibility (b) inverse charge susceptibility variation with temperatures for $\delta= 0.1$}
    \label{fig:sus}
\end{figure}

Figure \ref{characteristic_freq_scale} presents the characteristic frequency scales for spin and charge fluctuations across various doping levels, while Fig.\ref{fig:sus} depicts susceptibilities as a function of temperature, highlighting their significant influence on \( P_{n/s}(\omega) \). Our findings indicate that spin diffusion enters the classical regime at lower temperatures, suggesting that spin excitations shift to classical behavior sooner than charge excitations. Conversely, charge diffusion only transits to the classical regime at higher temperatures. Notably, charge excitations retain quantum properties even as they approach temperatures typically associated with the crossover to classical behavior, showing persistent quantum dynamics at the boundary between quantum-incoherent and classical regimes.

The observation that spin diffusion becomes classical at lower temperatures than charge diffusion is noteworthy. In strongly correlated systems, the decoupling of spin and charge dynamics—where spin diffusion turns classical while charge remains quantum—must signal anomalous transport properties. This decoupling could contribute to linear temperature-dependent resistivity, where charge carriers interact with a diffusive, classical spin background.  Put another way, strong inelastic scattering of fermions off dynamically fluctuating (partially unquenched) local moments will lead to a linear-in-$T$ resistivity.  
There will be a second linear-in-$T$ regime, where dissipation is further enhanced because fermions scatter off both, unquenched spin and charge pseudospin fluctuations.  We associate this latter regime with the bad metal.

The persistence of quantum behavior in charge excitations, even as local moments remain unquenched implies a ``decoupling'': charge transport is governed by quantum coherence while spin dynamics become anomalously diffusive. This interplay disrupts standard electron scattering processes, involving a two-stage restoration of quasiparticle coherence and leading to non-Fermi liquid behavior, such as linear resistivity.  True one fermion coherence now involves quenching of both, charge and spin fluctuations, and occurs only below $T_{coh}=T_{FL}$.

\FloatBarrier

\subsection{Kullback-Leibler Deviation Analysis}
Analyzing the diffusion spectrum \( P_{n/s}(\omega) \) through the Kullback-Leibler (KL) divergence provides deeper insights beyond those offered by the characteristic frequency scale \( \Omega_{n/s} \). While \( \Omega_{n/s} \) captures the energy scale at which quantum-to-classical transitions occur, the KL divergence highlights how the overall distribution shifts across temperature regimes, offering a more detailed picture of the system's evolution.

The KL divergence is particularly useful for identifying anomalous temperature ranges in the cdrossover regions that may not align neatly with the expected Fermi liquid or incoherent behavior. This analysis can reveal intermediate or mixed regimes that \( \Omega_{n/s} \) might overlook, providing an additional layer of understanding about the system's behavior. By measuring how \( P_{n/s}(\omega) \) departs from a reference state, such as a Fermi liquid distribution, the KL divergence can pinpoint subtle deviations that characterize the crossover from coherent to incoherent behavior.

\begin{figure}
    \centering
    \includegraphics[width=0.75\linewidth]{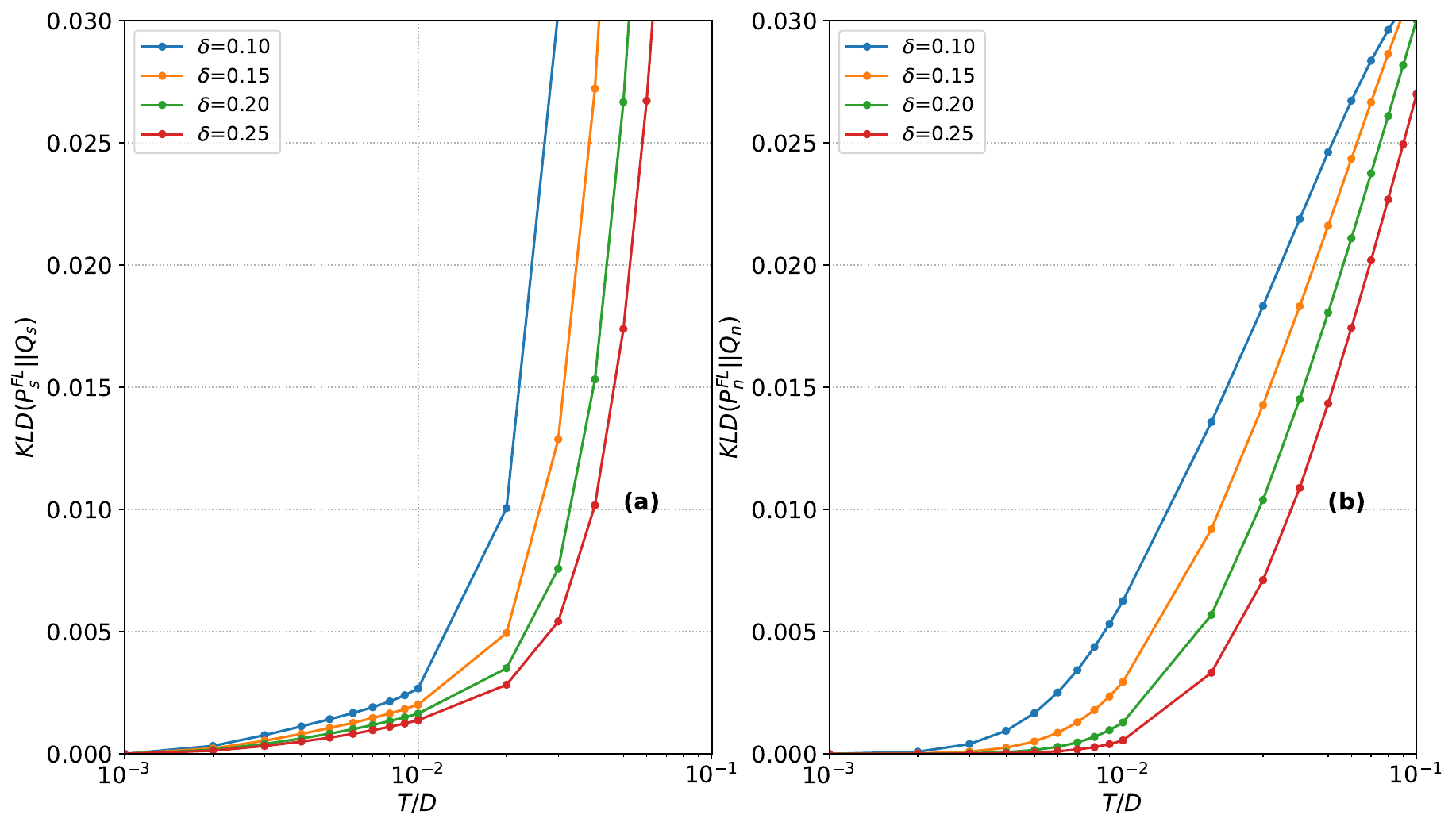}
    \caption{(a) KL divergence for spin diffusion spectra with reference distribution in FL phase.(b) KL divergence for charge diffusion spectra with reference distribution in FL phase. }
    \label{KLD_FL_ref}
\end{figure}

Furthermore, the KL divergence is sensitive to the finer features of the diffusion spectra, such as variations in width, shape, and peak positions, which may not be fully captured by \( \Omega_{n/s} \). This sensitivity allows for a more detailed assessment of how energy dissipation and transport properties evolve with temperature. For instance, differences in the way spin and charge spectra respond to temperature changes can be highlighted, shedding light on the underlying interactions and correlation effects that influence transport and scattering processes.
Given a reference probability distribution $P(\omega)$ , the KL Divergence with respect to another distribution $Q(\omega)$ is defined as 
\begin{align}
    KLD(P||Q) = \int_{-\infty}^{\infty} d\omega P(\omega) \log{\frac{P(\omega)}{Q(\omega)}}
\end{align}
Fig. \ref{KLD_FL_ref} illustrates the KL deviation for spin and charge when compared to their respective Fermi liquid reference distributions. The analysis shows that the spin KLD remains aligned with the Fermi liquid reference up to a crossover temperature lower than that for charge, indicating that charge excitations maintain coherent quasiparticle-like quantum behavior over a broader temperature range. This reflects the separate scales associated with persistence of quantum coherence in charge and spin dynamics before both eventually transit to a more classical, diffusive state.

In contrast, the charge KLD begins to deviate from the Fermi liquid reference at a higher temperature, suggesting that charge excitations lose their coherent, Fermi liquid character at $T$ higher than that for spin excitations.  This difference in the temperature behavior of spin and charge KLDs has significant implications for understanding transport properties in strongly correlated systems, indicating a quicker breakdown of quantum coherence for spin fluctuations as $T$ increases.

The KL deviation offers a complementary perspective to \( \Omega_{n/s} \) by providing a quantitative measure of how closely the system's state at a given $T$ matches an established regime, such as a Fermi liquid or an incoherent state. This added layer of analysis can help uncover subtle shifts and complex crossovers that the characteristic frequency scale alone might not reveal, thereby helping enhance our understanding of the intricate interplay between spin and charge dynamics in the doped Hubbard model.

\subsection{Understanding Kurtosis in the Diffusion Spectrum}
Examining the kurtosis of spin and charge diffusion spectra provides valuable insights that go beyond those offered by the characteristic frequency scale \( \Omega_{n/s} \) and the KL divergence. While \( \Omega_{n/s} \) and KL divergence offer a broad understanding of energy scales and deviations from reference distributions, kurtosis delves deeper into the shape and tail behavior of the spectrum, shedding light on the nature of excitations within the system.

Kurtosis, defined as:

\begin{align}
Kurt = \frac{\int_{-\infty}^{\infty} \omega^{4} P_{n/s}(\omega) d\omega}{\left(\int_{-\infty}^{\infty} \omega^{2} P_{n/s}(\omega) d\omega\right)^2},
\quad \text{Excess Kurtosis} = Kurt -3
\end{align}

quantifies the sharpness or ``tailedness" of a distribution. High kurtosis indicates a spectrum with pronounced peaks and heavy tails, signifying concentrated energy dissipation or fluctuations around specific frequencies. Conversely, low kurtosis suggests a flatter, more evenly distributed spectrum, pointing to a more uniform spread of excitations. This analysis provides a detailed view of the system's behavior across temperature changes, highlighting aspects that characteristic frequency scales and KL divergence do not capture.

While \( \Omega_{n/s} \) provides an average measure of excitation energy and the KL divergence helps identify deviations from reference states to illustrate how distributions shift away from known behaviors, neither metric addresses the distribution's sharpness or tailedness directly. This is where kurtosis brings additional insights. For example, high kurtosis at low temperatures indicates sharp, coherent excitations characteristic of a Fermi liquid state. As temperature increases, a decrease in kurtosis could imply a broadening of the distribution, signaling the shift to more diffuse excitations typical of an incoherent or classical state.

The kurtosis analysis also helps differentiate the temperature response of spin and charge excitations.  This complements findings from the KL divergence, indicating not only when spin and charge begin to diverge from Fermi liquid behavior but also how the distribution properties change as they move toward incoherence.

Overall, kurtosis complements the analysis by showing when the system retains sharp excitations or transitions to a more uniform distribution with temperature. This helps understanding the stability of spin and charge excitations as they move from quantum coherence to classical diffusion. When combined with insights from \( \Omega_{n/s} \) and the KL divergence, kurtosis offers a comprehensive view of how spin and charge dynamics evolve across temperature regimes.  Since kurtosis can be experimentally extracted from charge- and spin-fluctuation spectra and related to the corresponding dynamical susceptibilities, these connections could be testable.

\begin{figure}
    \subfloat[]{\includegraphics[width=0.45\linewidth]{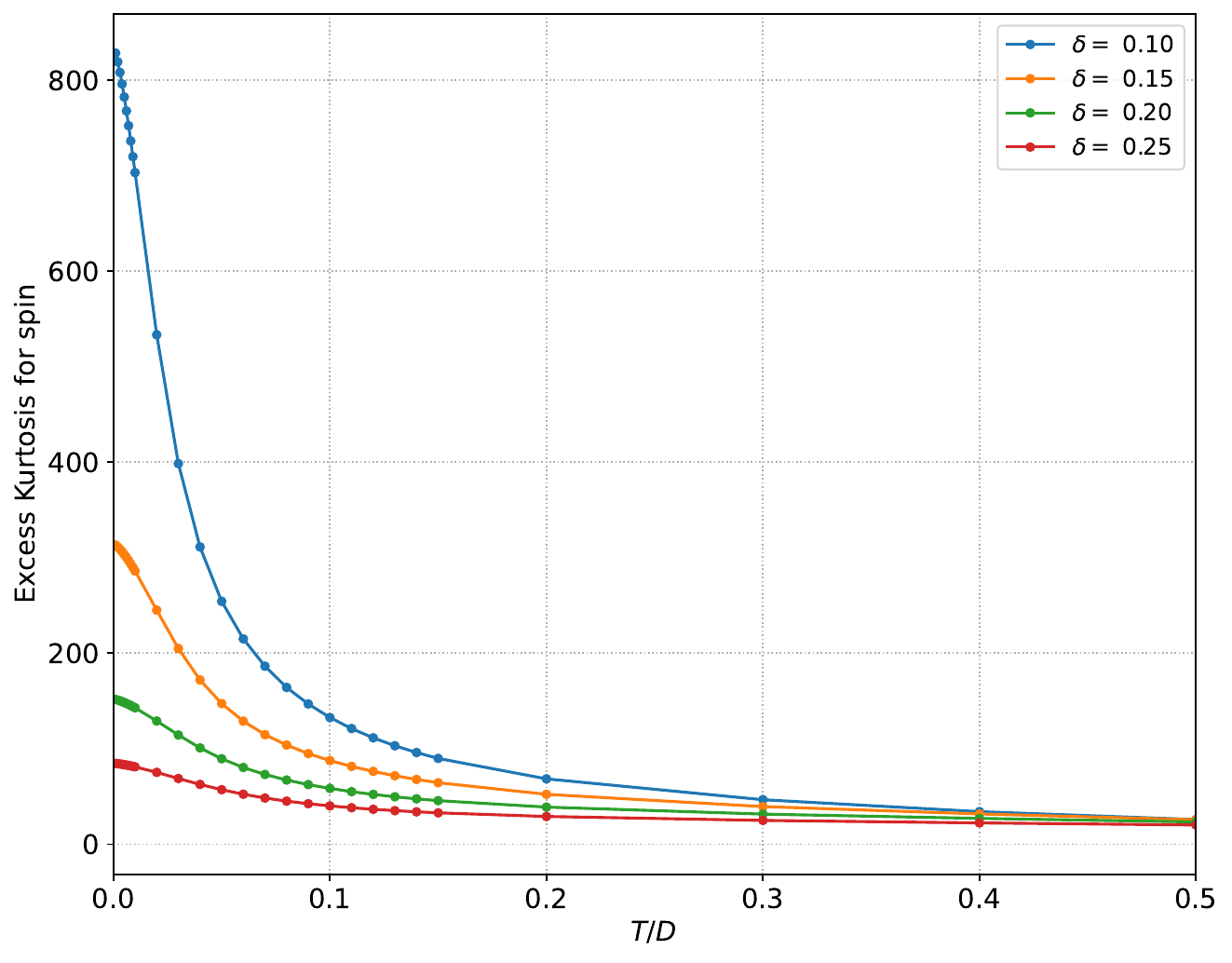}}
    \subfloat[]{\includegraphics[width=0.45\linewidth]{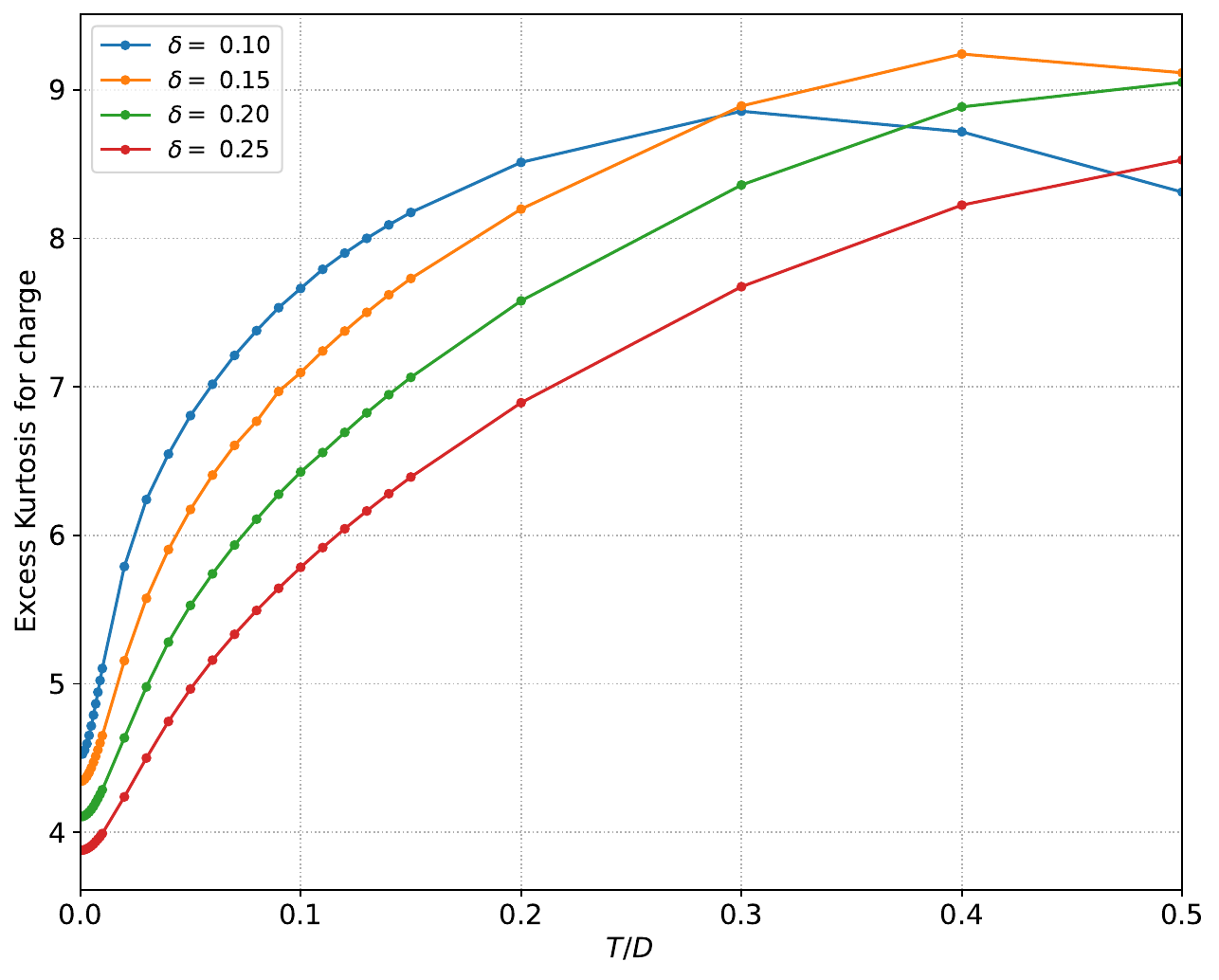}}
    
    \caption{Excess kurtosis for spin and charge diffusion spectra at $U=4$}
    \label{Excess_Kurt}
\end{figure}

Fig. \ref{Excess_Kurt} illustrates the kurtosis for spin and charge diffusion spectra. The results indicate that spin kurtosis is significantly higher at low temperatures, reflecting sharp, well-defined excitations typical of coherent Fermi liquid behavior. As temperature rises, spin kurtosis decreases, indicating a broadening of the distribution as the system transitions to a more classical or incoherent state. Conversely, the kurtosis for charge begins at a separate temperature, implying more evenly distributed, less sharply defined 
excitations.  This accords with indications from other measures above, and supports a two-stage recovery of quasiparticle coherence. 

Finally, charge kurtosis is around 100 times smaller than spin kurtosis, further underscoring the distinct nature of spin and charge dynamics. It shows that spin excitations dominate and are more prone to maintaining coherent, sharp features at low temperatures, while charge excitations demonstrate more gradual changes and broader distributions as temperature varies. In view of this, it is now quite remarkable that the system manages to achieve full quantum coherence (below $T_{FL}$) at all.  Our choice of measures, we believe, go a meaningful way toward providing some understanding of this process.

\subsection{Inverse Diffusion Constant Analysis}

\begin{figure}[h]
    \includegraphics[width=0.5\linewidth]{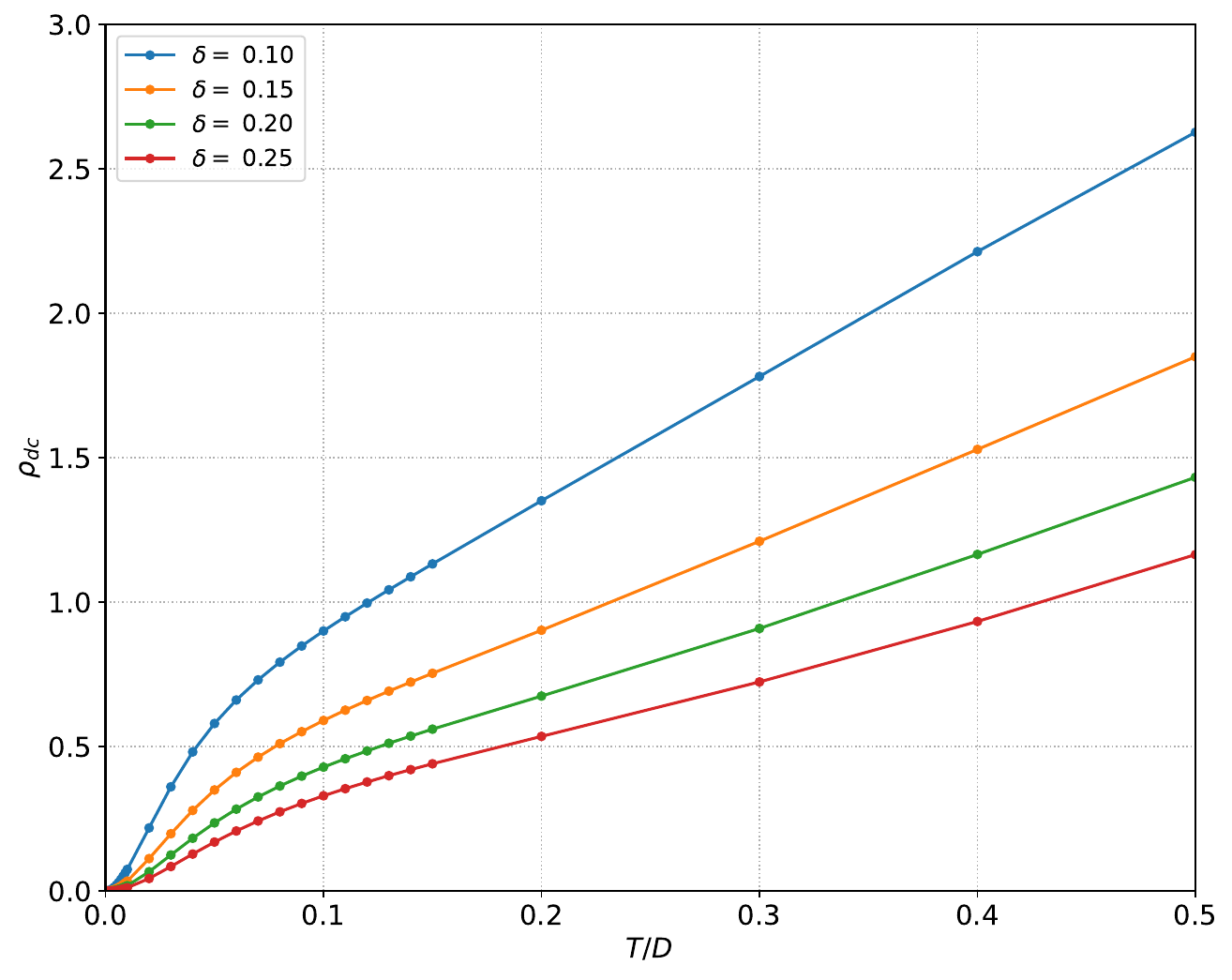}
    \caption{Variation of resistivity with Temperature for various doping values at $U=4$}
    \label{Resistivity_plot}
\end{figure}

The diffusion constants for charge and spin are given by \cite{10.1063/1.1456431}:

\begin{align}
    D_{n} = \frac{\sigma(0)}{\chi_{n}}, \quad D_{s} = \frac{\sigma(0)}{4\chi_{s}}
\end{align}

where \(\sigma(0)\) represents the dc conductivity. The temperature-dependent resistivity, defined as \(\rho_{dc} = \frac{1}{\sigma(0)}\), is shown for various doping levels in Fig.\ref{Resistivity_plot}.

Examining the inverse of the diffusion constants, \( D_{n/s}^{-1} \), derived from transport properties bridges the microscopic features of the diffusion spectrum \( P_{n/s}(\omega) \) with the macroscopic behavior of the system. While the characteristic frequency scale \( \Omega_{n/s} \), KLD, and kurtosis provide insights into the energy scales, coherence, and distribution properties of the spectrum, \( D^{-1} \) offers a more direct measure of transport phenomena such as the resistivity.

\begin{figure}[h]
    \centering
    \includegraphics[width=0.75\linewidth]{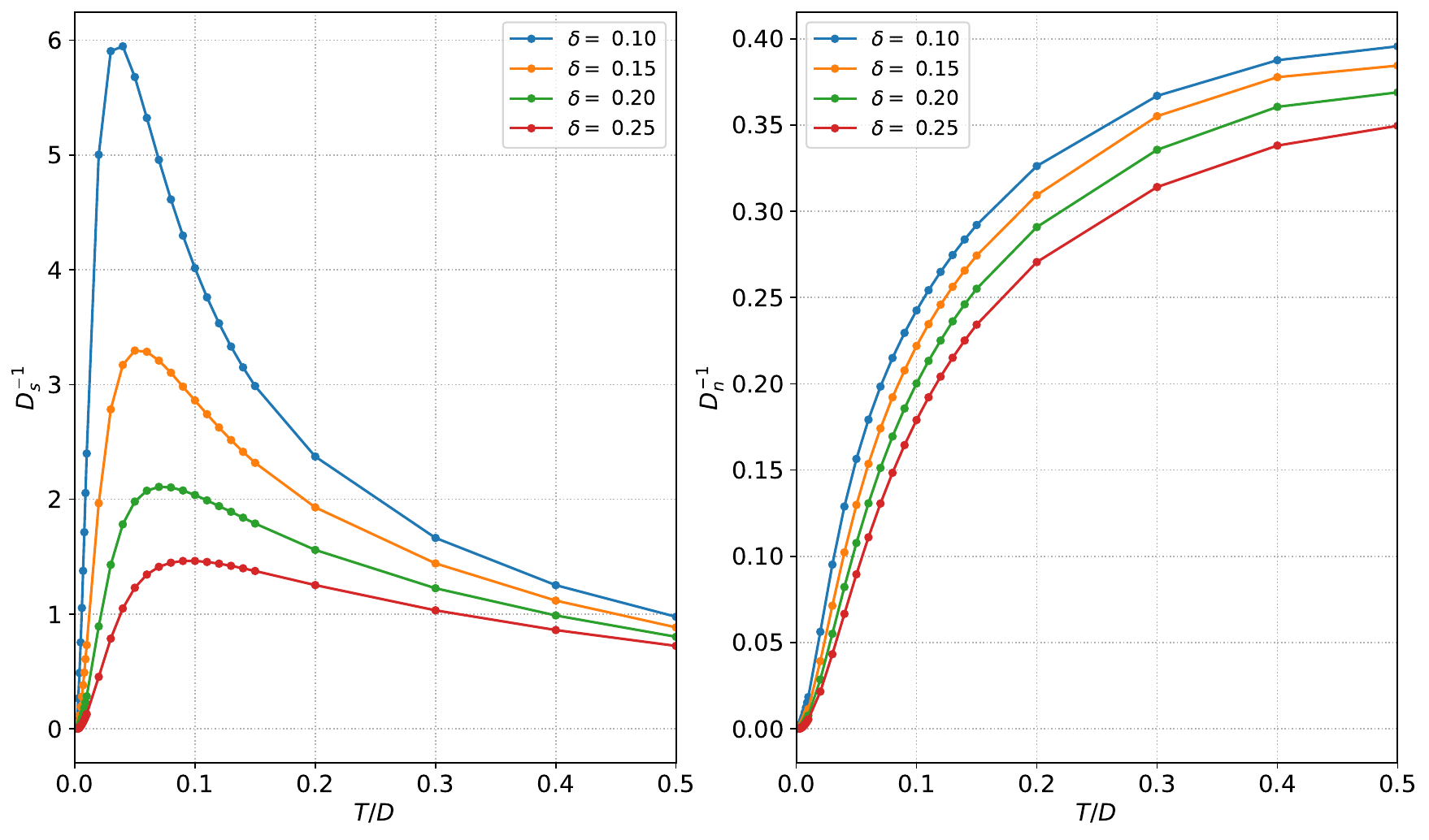}
    \caption{Inverse of spin and charge diffusion constant at $U=4$.}
    \label{Inverse_diff_const}
\end{figure}
In Fig. \ref{Inverse_diff_const}, we illustrate the temperature dependence of the inverse diffusion constant for both spin and charge. The behavior for spin shows an initial increase as temperature rises, indicating a growing resistance to transport. This increase is followed by a decrease and eventual saturation at higher temperatures, suggesting a stabilization of spin transport in a more classical regime. In contrast, the inverse diffusion constant for charge consistently increases with temperature before plateauing, implying an increasing resistance that levels off at higher temperatures. Notably, the inverse diffusion constant for charge is about ten times lower than that for spin across the temperature range, highlighting a fundamental difference in transport behavior between the two.

These observations underscore the distinct dynamics governing spin and charge transport. The distinct temperature dependencies suggest that spin excitations may be more prone to coherence loss and scattering, transitioning through regimes of varying transport resistance. In comparison, charge excitations show a smoother shift towards saturation, maintaining a higher degree of mobility despite temperature changes. This disparity aligns with the findings from kurtosis and KLD analyses.

The diffusion constant \( D \) itself represents how freely spin or charge moves through the system, linking directly to transport efficiency. The spectrum \( P_{n/s}(\omega) \) encapsulates energy dissipation and fluctuation behavior, revealing the scattering mechanisms that influence \( D \). By exploring \( D^{-1} \), we can discern how the specific features of \( P_{n/s}(\omega) \)—such as peaks, width, and tails—contribute to transport. High kurtosis in \( P_{n/s}(\omega) \) suggests concentrated energy dissipation around certain frequencies, leading to reduced scattering and higher mobility (lower \( D^{-1} \)). A broader spectrum, by contrast, implies increased scattering and diffusion, resulting in a higher \( D^{-1} \) and decreased transport efficiency.

The analysis of \( D^{-1} \) in relation to \( P_{n/s}(\omega) \) provides further evidence of the decoupling between spin and charge transport. For example, if spin \( D^{-1} \) increases more significantly than that of charge with temperature, it signals that spin excitations transit to an incoherent state at a rate distinct from their charge counterparts. This complements kurtosis and KLD findings, which indicate how spin and charge fluctuations achieve quantum coherence at separate scales. This difference is critical for understanding the non-Fermi liquid behavior and it's eventual crossover to a Landau FL, often observed in strongly correlated systems.

This plays a crucial role in interpreting \( D^{-1} \) alongside \( P_{n/s}(\omega) \), KLD, and kurtosis. As temperature increases, a rising \( D^{-1} \) paired with a decreasing kurtosis suggests a transition from sharp, coherent quasiparticle excitations to incoherently diffusing modes. Thus, taken together, these measures offer a comprehensive characterization of how dynamic charge and spin fluctuations evolve and shape the transport properties across the Mott 
transition in the local approximation for the one-band Hubbard model.

\section{Conclusion}
\label{conclusion}
In this study, we have conducted detailed analysis of spin and charge transport properties for the doped Hubbard model, employing both Density Matrix Numerical Renormalization Group (DMNRG) and Full Density Matrix (FDM) methods. The comparative examination of these methods revealed their respective strengths and limitations, particularly in capturing low-frequency behavior in diffusion spectra. DMNRG proved superior for low-temperature analyses, accurately resolving the sharp, coherent features indicative of strong correlations, while FDM, though valuable for broader temperature ranges, exhibited numerical artifacts such as spurious dips at low frequencies. 

Key metrics such as the characteristic frequency scale, Kullback-Leibler divergence (KLD), kurtosis, and the inverse diffusion constant provided complementary insights into the spin and charge dynamics across the bandwidth and filling-driven Mott transitions, asd well as across the coherence-incoherence crossover. Our analysis highlighted that spin and charge excitations exhibit distinct temperature dependencies, with spin transport transitioning to classical behavior at lower temperatures compared to charge. This decoupling between spin and charge dynamics underpins multiscale crossovers in transport. The persistence of quantum characteristics in charge excitations at higher temperatures, even as spin behavior becomes more classical, controls the nature of the $T$-dependent crossovers in resistivity from quantum-coherent, via quantum-incoherent, into the ``classical'' regimes.

Overall, the findings illustrate that combining analyses from various metrics offers a comprehensive understanding of the transition from coherent to incoherent transport. The insights gained from the distinct behavior of spin and charge excitations, their response to temperature, and the interplay between quantum and classical regimes contribute to a deeper understanding of non-Fermi liquid behavior and transport anomalies in strongly correlated materials. This work emphasizes the importance of using multiscale approaches to unearth the full complexity of spin and charge transport in correlated systems, and suggests avenues for refining computational methods to overcome numerical challenges in low-frequency regimes.

Finally, what about models where non-Fermi liquidity may persist down (hypothetically) to $T=0$?  While this is not possible in the one-band Hubbard model in the local approximation, it seems to obtain in
symmetry-unbroken metallic states in cluster extensions or in multi-orbital models, but only as long as a momentum- or an orbital-selective Mott phase obtains.  In such ``two-fluid'' phases, momentum- or orbital-selective persistence of the Mott insulator features with it's unquenched local moments (in the absence of any possibility of any ordering) implies
persistence of the incoherent charge and spin fluctuations down to $T=0$, simply because once a selective Mott transition occurs, the
local moments of the selectively Mott localized sector cannot be ``Kondo'' quenched by the one-electron hybridization with carriers in the metallic fermion sector.  Indeed, the one-electron hybridization is known to scale to zero in the OSMP~\cite{Delft}.  Guided by our analysis, one should thus expect absence of the low-$T$ quasicoherent regime in the various responses considered in this paper in any selective-Mott phases.  Since this is tied to the destruction of lattice Kondo screening in such phases, the low-energy incoherence, along with emergent infra-red branch-point analytic structure with fractional power-law decay in one- and two-fermion responses~\cite{Laad} could be attributable to the Anderson-Nozieres orthogonality catastrophe: after all, this is expected to immediately become operative when {\it coherent} one electron hybridization scales to zero in such models.  These phases would then resemble the quantum-incoherent and bad-metal incoherent regimes (discussed here) in the one-band Hubbard model above the Landau FL coherence scale.  Whether such a fluctuating state is a ``strange metal'' remains an open issue of considerable current interest.  It would be an interesting avenue to extend this work for such cases in future.

%#############################################################################################################################
\textbf{Acknowledgment}: Our sincere thanks go to H.R. Krishnamurthy for insightful and invaluable discussions. TVR is grateful for the support received from the DST384 as a Year of Science Chair Professor during the initial phase of this work. SRH appreciates the hospitality and access385
to facilities provided by JNCASR during visits.

\bibliography{hf}

\appendix
\section{Evaluating Low-Frequency Behavior with Cumulative Distribution Functions}
\label{app_A}
To quantify the accuracy of low-frequency behavior in the charge and spin diffusion spectra, we analyze the cumulative distribution function \( C(\omega) \):

\begin{align}
    C(\omega) = \int_{0}^{\omega} d \omega' \, \mathrm{Im} \chi(\omega'),
\end{align}

where \( \mathrm{Im} \chi(\omega) \) is the imaginary part of the susceptibility for both charge and spin. In a metallic phase, characterized as a Fermi liquid, \( \mathrm{Im} \chi(\omega) \) is expected to display a linear behavior at low frequencies, resulting in a log-log plot slope of \( C(\omega) \) approaching 2, indicative of a quadratic dependence.

\begin{figure}
    \centering
    \subfloat[]{\includegraphics[width=0.5\linewidth]{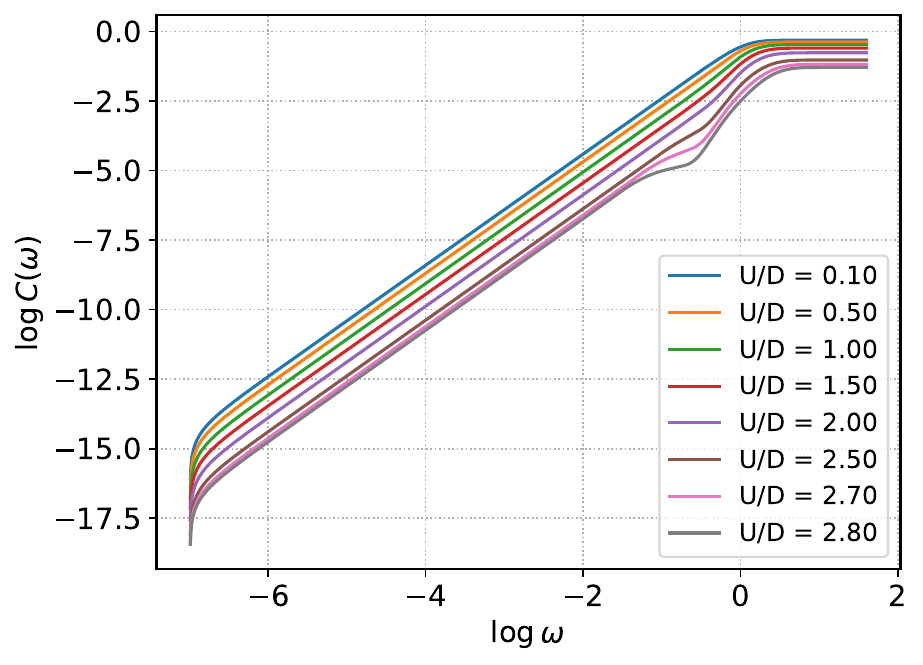}}
    \subfloat[]{\includegraphics[width=0.5\linewidth]{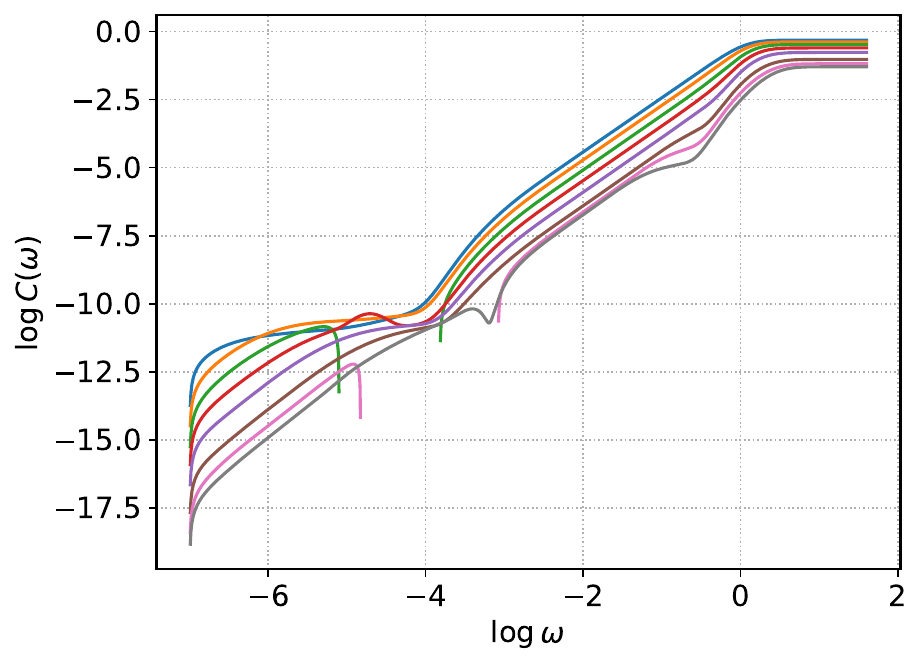}}
    \caption{(a) Cumulative distribution function \( C(\omega) \) for the charge diffusion spectrum at half-filling using DMNRG at $T/D = 0.001$. (b) Cumulative distribution function for the charge diffusion spectrum using FDM at $T/D = 0.001$.}
    \label{charge_cumulant_hf}
\end{figure}

\begin{figure}[h]
    \centering
    \includegraphics[width=0.5\linewidth]{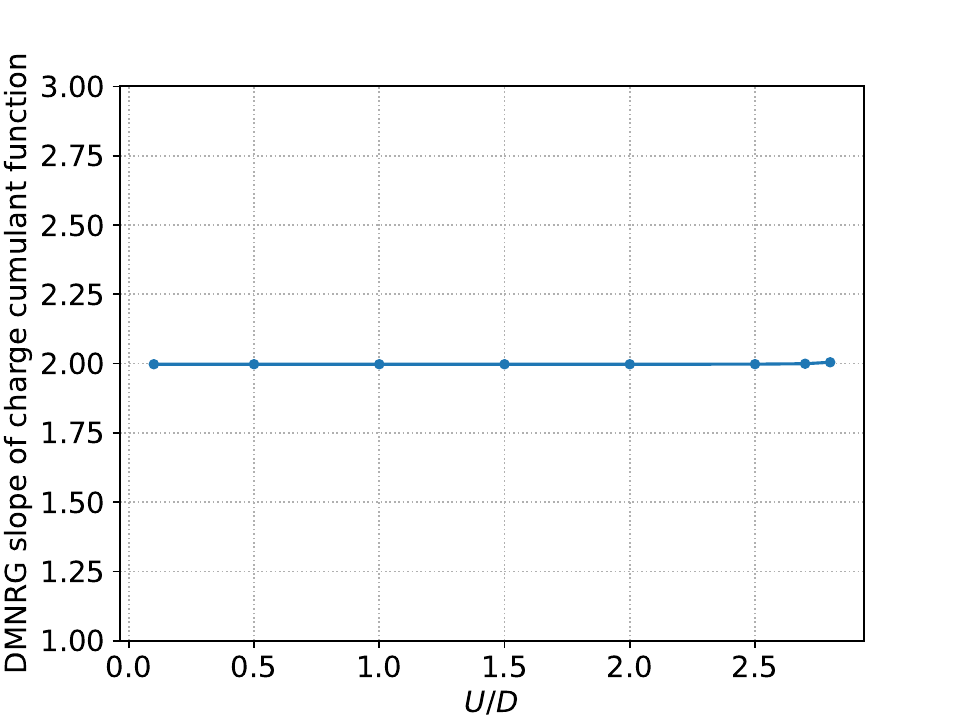}
    \caption{Fitted slope of $\log{C(\omega)}$ for the charge cumulant function for $\log{\omega}$ between \(-6\) and \(-2\) for DMNRG data, illustrating the consistency of the slope across various \( U \) values in the metallic phase.}
    \label{charge_slope_hf}
\end{figure}

\begin{figure}
    \centering
    \includegraphics[width=0.6\linewidth]{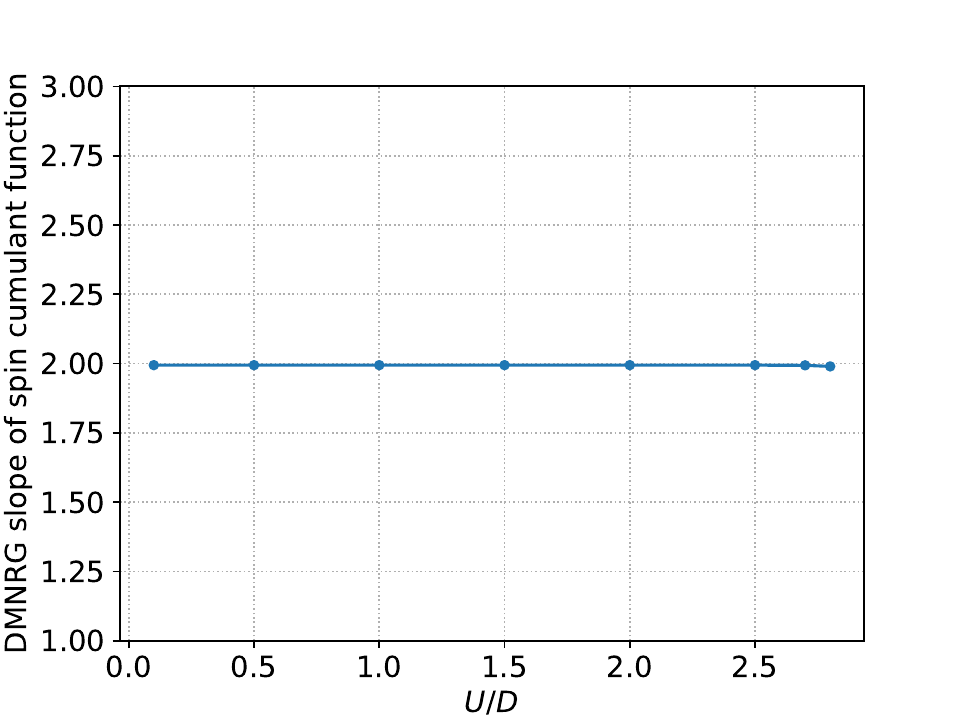}
    \caption{Fitted slope of $\log{C(\omega)}$ for the spin cumulant function for $\log{\omega}$ between \(-6\) and \(-2\) for DMNRG data, illustrating the consistency of the slope across various \( U \) values in the metallic phase..}
    \label{spin_slope_hf}
\end{figure}

The DMNRG results for \( C(\omega) \), shown in Fig. \ref{charge_cumulant_hf}(a), confirm the expected behavior, maintaining a low-frequency slope close to 2, consistent down to \( 10^{-6} \). The slope analysis in Figs. \ref{charge_slope_hf} and \ref{spin_slope_hf} corroborates this, showing consistency across various \( U \) values in the metallic phase. By contrast, the FDM results in Fig. \ref{charge_cumulant_hf}(b) show deviations at low frequencies, mirroring the numerical artofacts found above.

\begin{figure}
    \centering
    \subfloat[]{\includegraphics[width=0.5\linewidth]{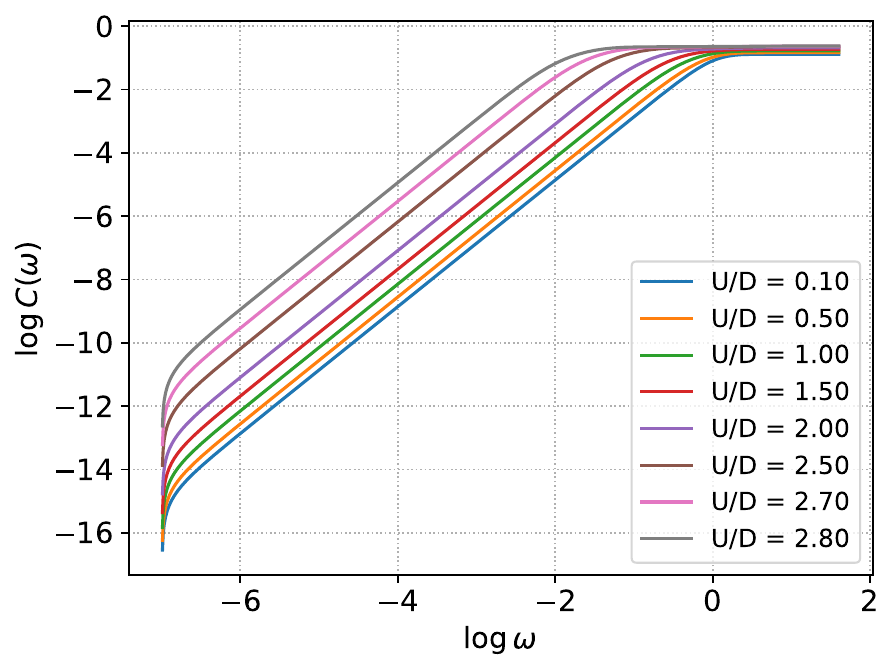}}
    \subfloat[]{\includegraphics[width=0.5\linewidth]{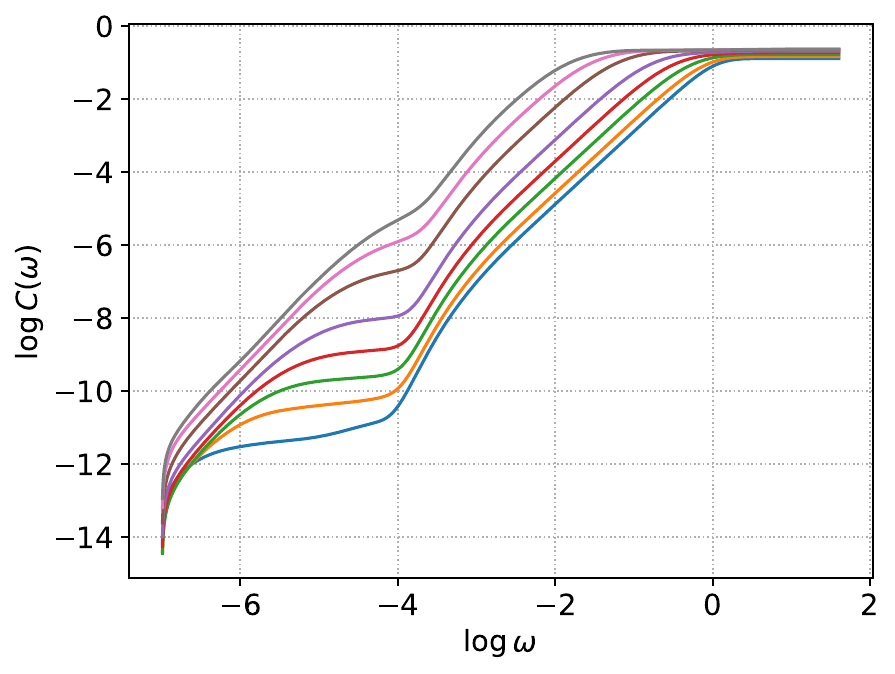}}
    \caption{(a) Cumulative distribution function \( C(\omega) \) for the spin diffusion spectrum at half-filling using DMNRG at $T/D = 0.001$. (b) Cumulative distribution function for the spin diffusion spectrum using FDM at $T/D = 0.001$.}
    \label{spin_cumulant_hf}
\end{figure}

Similar result for spin, depicted in Fig. \ref{spin_cumulant_hf}, reinforces these findings. The DMNRG data in Fig. \ref{spin_cumulant_hf}(a) uphold the expected low-frequency behavior, while FDM results in Fig. \ref{spin_cumulant_hf}(b) exhibit deviations, further highlighting the limitations of FDM.

The consistency of these results confirms DMNRG’s superior accuracy in capturing low-frequency spectral behavior, while FDM exhibits numerical artifacts that affect the slope and, consequently, the diffusion spectrum.

\section{Optimizing the Diffusion Spectrum: Mitigating Artifacts in the FDM Approach}
\label{app_B}
\begin{figure}[h]
    \centering
    \includegraphics[width=0.9\textwidth]{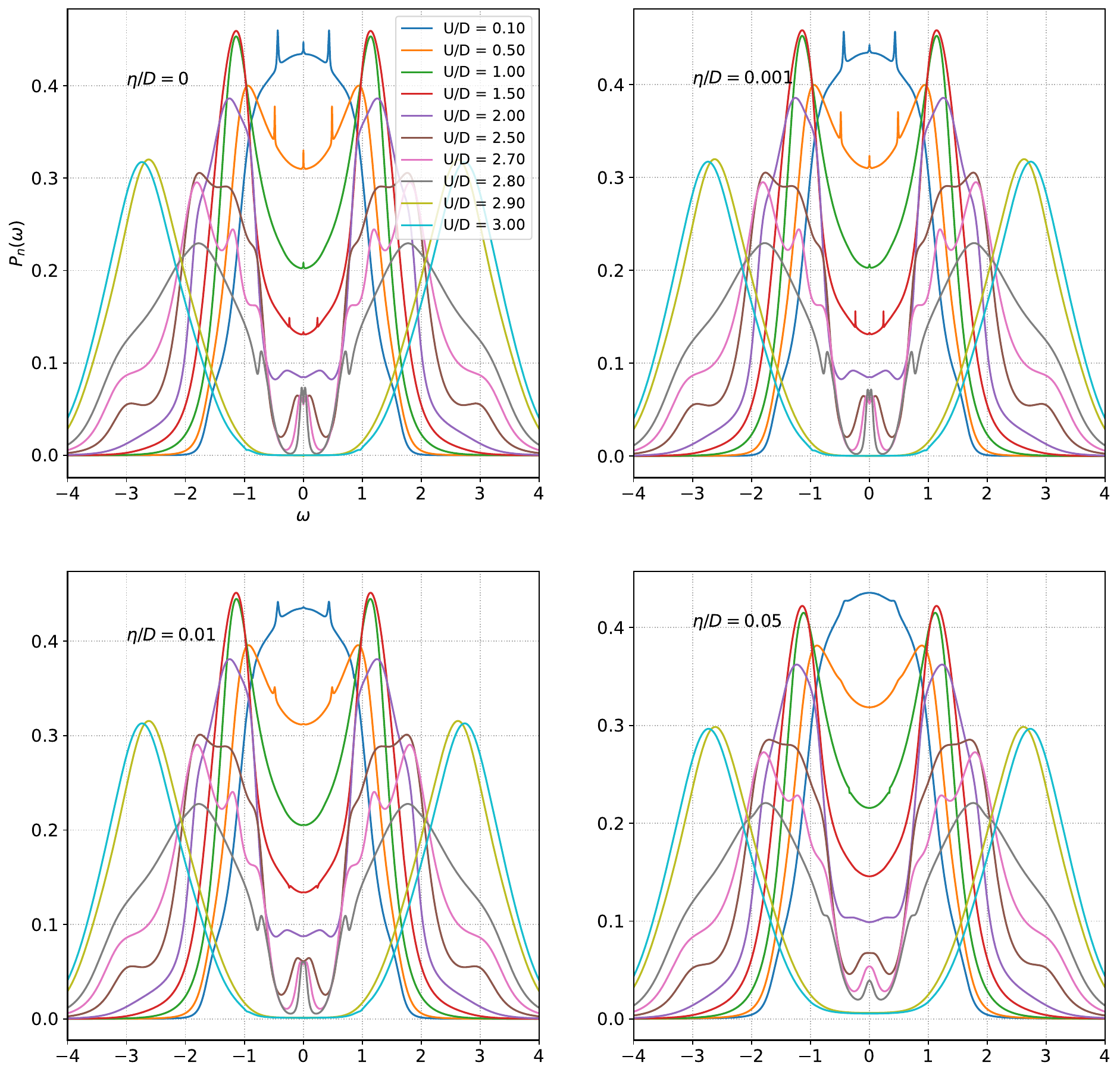}
    \caption{Charge diffusion spectrum using Pade analytic continuation for four different values of \(\eta\) at half-filling at $T/D=0.001$.}
    \label{charge_ac_hf}
\end{figure}

To deepen our understanding of the diffusion spectra and the numerical challenges associated with the Full Density Matrix (FDM) method, we closely examine how the results evolve when calculated across various parameter schemes, We now focus on mitigating the spurious dip observed in FDM results. Specifically, we analyze the charge diffusion spectrum at \(\eta = 0\), as illustrated in Fig. \ref{chargediff_hf}(b). A pronounced dip near \(\omega = 0\) is evident, which does not appear in the DMNRG-derived results.  We vary the parameter \(\eta\) to observe its effect on the charge diffusion spectrum. Fig. \ref{charge_ac_hf} demonstrates the impact of increasing \(\eta\) on the spurious dip. As \(\eta\) increases, the dip diminishes, indicating that adjusting \(\eta\) can minimize numerical artifacts. Our analysis suggests that an optimal \(\eta\) value of 0.001 strikes a balance, effectively removing the dip while maintaining the essential characteristics of the spectrum. This optimized result aligns closely with the DMNRG results for \(\eta = 0\), demonstrating that careful tuning of \(\eta\) can mitigate artifacts without compromising the physical accuracy of the spectrum.

\begin{figure}[h]
    \centering
    \includegraphics[width=0.9\textwidth]{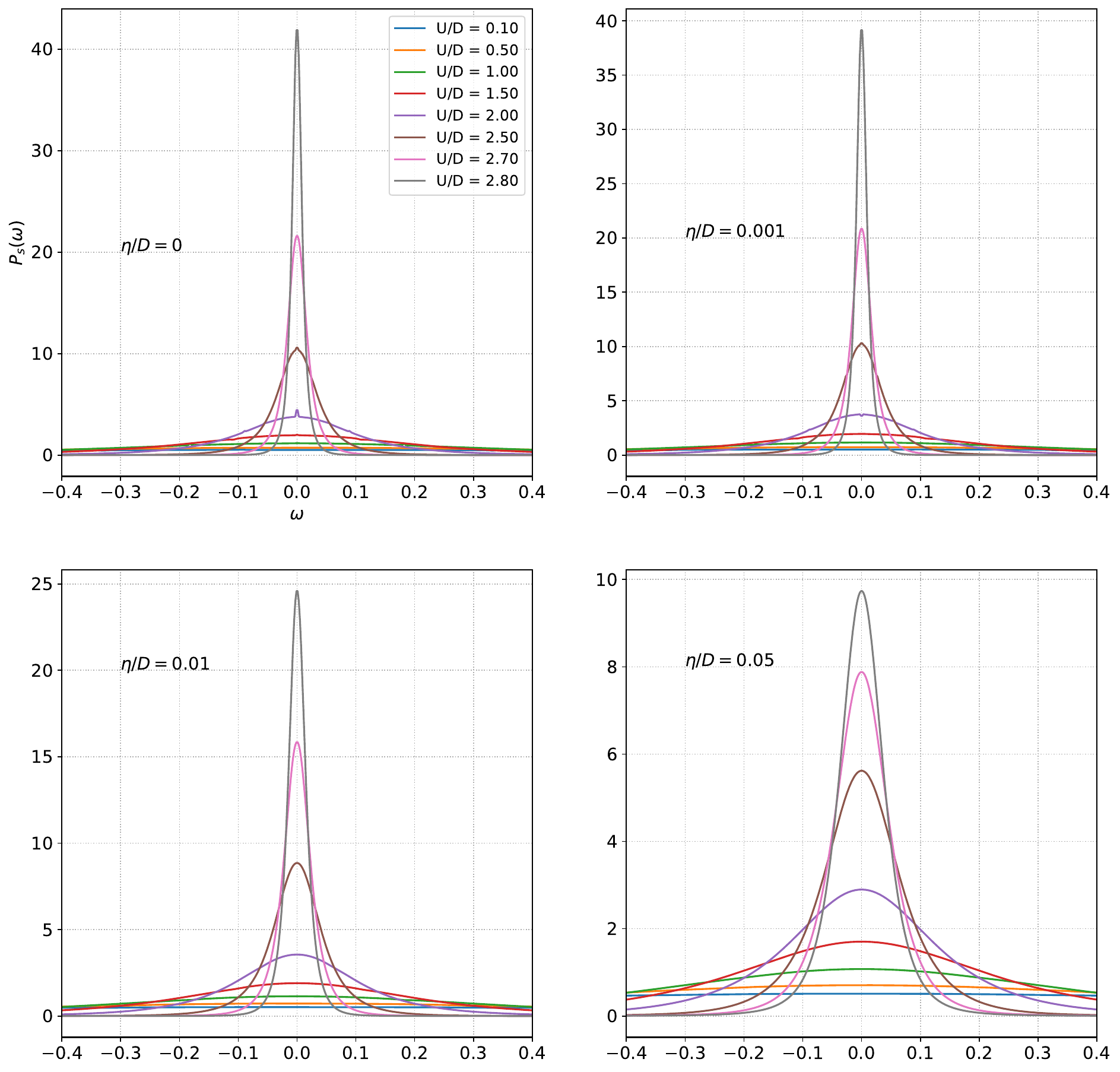}
    \caption{Spin diffusion spectrum using Pade analytic continuation for four different values of \(\eta\) at half-filling at $T/D=0.001$.}
    \label{spin_ac_hf}
\end{figure}

The analysis extends to the spin diffusion spectrum. Unlike the charge spectrum, the spin spectrum's spurious dip can be corrected with smaller adjustments to \(\eta\), as seen in Fig. \ref{spin_ac_hf}. Thus, charge response calculations in FDM are more sensitive to numerical artifacts than their spin counterparts, reflecting differences in how these channels respond to the same numerical treatments.

Figs. \ref{chargediff_hf}(b) and \ref{spin_cumulant_hf}(b) reinforce that while DMNRG accurately captures the expected diffusion spectra without artificial dips, the FDM approach faces challenges, particularly for \(\omega < T\) where quantum and thermal fluctuations dominate. This limitation is due to FDM's inherent difficulties in precisely representing the slope of the susceptibility at small frequencies, leading to artifacts in the diffusion spectrum.

Further examination using data obtained through analytical continuation (Figs. \ref{charge_ac_hf} and \ref{spin_ac_hf}) provides additional clarification. Even with very small broadening values \(\eta\), the analytically continued data does not display the spurious dips seen in the FDM results, reinforcing the conclusion that these dips are numerical artifacts rather than physical phenomena.

\end{document}